\documentclass{ws-ijmpa}

\usepackage{amssymb}
\usepackage{epsfig}
\usepackage{graphicx,psfrag}
\usepackage{bm}
\newcommand{\nn}{\nonumber \\}

\def\L{\mathcal{L}}
\def\D{\mathcal{D}}
\def\R{\mathbb{R}}
\def\N{\mathbb{N}}
\def\d{\partial}

\def\le{\left}
\def\ri{\right}
\def\phic{\phi_{cl}}
\def\vphi{\varphi}

\def\be{\begin{equation}}
\def\ee{\end{equation}}
\def\bei{\begin{itemize}}
\def\eei{\end{itemize}}
\def\beq{\begin{eqnarray}}
\def\eeq{\end{eqnarray}}
\def\vphi{\varphi}
\def\tu{\tilde{u}}
\def\tv{\tilde{V}}
\def\ts{\tilde{S}}
\def\tp{\tilde{\phi}}
\def\tb{\tilde{\beta}}

\begin{document}

\markboth{V. Pangon}
{Structure of the broken phase of the sine-Gordon model using functional renormalization}

%
\catchline{}{}{}{}{}
%

\title{Structure of the broken phase of the sine-Gordon model using functional renormalization
}

\author{V. Pangon}

\address{Gesellschaft f\"ur Schwerionenforschung mbH, Planckstr. 1\\
Darmstadt, D-64291,Germany\\
v.pangon@gsi.de}
\address{Frankfurt Institute for Advanced Studies, Universit\"at Frankfurt,\\
Frankfurt am Main, D-60438, Germany}
\address{{\it on leave to :} DiscInNet Labs, 5 rue de l'Eglise, 92700 Boulogne, France }

\maketitle
\begin{history}
\received{Day Month Year}
\revised{Day Month Year}
\end{history}

\begin{abstract} 
We study in this paper the sine-Gordon model using functional Renormalization Group (fRG) at Local Potential Approximation (LPA) using different RG schemes. In $d=2$, using Wegner-Houghton RG we demonstrate that the location of the phase boundary is entirely driven by the relative position to the Coleman fixed point even for strongly coupled bare theories. We show the existence of a set of IR fixed points in the broken phase that are reached independently of the bare coupling. The bad convergence of the Fourier series in the broken phase is discussed and we demonstrate that these fixed-points can be found only using a global resolution of the effective potential. We then introduce the methodology for the use of Average action method where the regulator breaks periodicity and show that it provides the same conclusions for various regulators. The behavior of the model is then discussed in $d\ne 2$ and the absence of the previous fixed points is interpreted.
\end{abstract}

\ccode{PACS numbers: 05.10.Cc,64.60.ae,12.38.Lg,11.10.Gh,11.30.Qc,11.10.Kk}

\section{Introduction}

The sine-Gordon model and its phase diagram has received a considerable interest these last decades due to its unique properties. 
In the euclidean spacetime, its Lagrangian reads :
\be\label{bare}
\L=\frac{1}{2}(\d_\mu\phi)^2+u\, cos(\beta\phi)
\ee

Most of our knowledge has been established in two dimensions. It is has been shown perturbatively -for $u<<1$- to exhibit two phases separated by a Kosterlitz-Thouless transition \cite{Kosterlitz:1973xp}. The spontaneously broken periodicity phase for $\beta<\beta_c$ with asymptotic freedom and a non-renormalizable one for $\beta>\beta_c$ separated by the Coleman frequency $\beta_c=\sqrt{8\pi}$ \cite{Coleman:1974bu}. 

The spontaneously broken phase is of major interest as it has been demonstrated to be equivalent to the neutral sector of the massive Thirring model at zero \cite{Coleman:1974bu,Mandelstam:1975hb} and finite temperature \cite{Delepine:1997bz}. At the partition function level, the sine-Gordon model can also be mapped onto the neutral Coulomb gas \cite{Samuel:1977vy,Amit:1979ab}. The sine-Gordon model has also been demonstrated to be dual to the $XY$ model in the Villain approximation \cite{Villain:1974ir} for vanishing vortex fugacity \cite{Huang:1990via}. And finally its $S$-matrix has been exactly computed in \cite{Arefeva:1974bk} as well as the $S$-matrix for its solitons \cite{Zamolodchikov:1978xm}.

In higher dimensions, the mapping at the partition function level between sine-Gordon and the Coulomb gas still holds as well as its relation to the non-linear sigma model in the small explicit breaking limit \cite{Samuel:1977vy} but little is known on these models, especially in the IR of their broken phase where non-perturbative techniques use is mandatory \cite{Kosterlitz77}. 

In addition to the previous relations with other models, the sine-Gordon model is also the simplest non-trivial theory with a field compact variable, a feature shared with non-abelian gauge theories. In particular, it is believed to be an effective theory for the deconfinement transition of $SU(2)$ \cite{Johnson:1991qc,Polonyi:1995jz}. The phenomenological viability of this approach will be fully addressed in another paper \cite{Pangon:2010ha}. In any case, it is also commonly accepted that the computation of the effective potential for the Polyakov-loop requires the study of periodic scalar field theories, see e. g. \cite{Weiss:1980rj}. Let us finally mention that direct studies of Yang-Mills theories using functional methods made recently major breakthrough, e. g. \cite{Zwanziger:2001kw,Lerche:2002ep,Pawlowski:2003hq,Fischer:2006ub,Fischer:2006vf,Fischer:2008uz,Marhauser:2008fz,Braun:2010cy}. We will show below that our findings comfort these approaches.

Let us now briefly introduce the challenges encountered when studying the broken phase of the sine-Gordon model. 
While the UV of the broken phase is well under control being perturbatively renormalizable, the negativity of the $\beta$-function for the amplitude of the model tells us that the IR will be strongly coupled, making the use of non-perturbative techniques such as functional renormalization mandatory (for reviews, see \cite{Polonyi:2001se,Bagnuls2001,Berges2002,Pawlowski:2005xe,Gies:2006wv,Delamotte:2007pf,Rosten:2010vm}). This feature has been studied using functional renormalization group in $d=2$ \cite{Nandori:1999vi,Nandori:2005pa,Nagy:2006se,Nagy:2006pq,Nagy:2006ue,Nagy:2009pj,Nandori:2009ad,Kehrein:1999nx} where a possible singularity related to the convexity limit of the functional flows tends to appear in the IR. This same feature was also conjectured in $d\geq3$ \cite{Nandori:2003pk}. 
The methodology of these studies consists in following the flow for the Fourier coefficients of the effective action, thus  implicitly assuming that the relevance classification in the UV with respect to the Coleman fixed point still holds at least partially in the IR. We will detail why this assumption is incorrect in $d=2$ due to the appareance of novel fixed points that can not be found using Fourier series, in agreement with previous findings \cite{Pangon:2009wk}. The possible instability of the flow in $d>2$ will also be shown to be a genuine one and thus of a different nature.

While we believe that the non-pertubative behavior of the sine-Gordon model at LPA is now fully understood with the new results provided here, the inclusion of the wave function renormalization beyond the inclusion of a field-independent anomalous dimension \cite{Nagy:2009pj}  at the non-perturbative level is still missing. This point is obviously of crucial importance in the case the sine-Gordon model as one expects a non-trivial anomalous dimension to yield the Kosterlitz-Thouless cross-over already at the perturbative level. 
The difficulty lies in the fact that the Effective Average Action (EEA) approach \cite{Wetterich:1989xg,Wetterich:1992yh} which allows to define properly the next to leading order in the gradient expansion is based on the addition of a regulator that breaks explicitly the periodicity. As a result, the use of this RG scheme is non-trivial and some justifications have to be worked out in the case of the sine-Gordon model.
On the other hand, the RG scheme of Wegner-Houghton (WH) \cite{Wegner1973} does not suffer from this plague as the fluctuations are computed explicitly with a loop expansion for some small parameter (see below). Its sharp cut-off nature does not allow the consistent definition of the wave-function renormalization but can provide at LPA  reliable results that could be used as a reference for the EEA computations.

\subsection{Outline}
As a result from the previous considerations, we will adopt the following strategy. 

In section II, we introduce WH-RG and briefly discuss the main properties of its flow equation.
We focus on the case $d=2$ and point out the relevant quantities that characterize the flow as one expects the effective potential to be both convex and periodic at k=0 i. e. constant in both phases. 
We study perturbatively the location of the phase boundary and show the presence of a continuum of non-perturbative fixed points in the IR of the broken phase  previously found in \cite{Pangon:2009wk}. We then discuss the convergence property of the Fourier representation of the potential and show that only a global-non-Fourier-expanded numerical treatment can describe accurately the broken phase and prevent the appearance of singularities in the IR flow. The set of fixed points, which stabilizes the loop expansion and ensures a safe convex limit for $k\to 0$ is then shown to be reached independently of the bare coupling $u$. We also extend the validity of the phase boundary to strongly coupled bare theories and demonstrate that the renormalized potential depends only on $\beta$.

In the section III, we derive the average action flow equation and project it on the LPA in $d=2$. The evolution equation for the potential is then discussed and the impact of the regulator on the location of the perturbatively-established phase boundary is discussed. In particular, we demonstrate that all the regulators provide the same phase boundary. The Fourier series convergence is also discussed and is shown to be even worse than in WH's case. It is then shown that for various regulators, all the features found in the previous section still hold.

In section IV, we introduce the main features of the sine-Gordon model in $d\ne2$. The UV perturbative scaling regime is exhibited and the phase diagram is discussed. The IR fixed points found in $d=2$ are shown to diseapear and a fundamental instability of the flow in the IR for $d>2$ is emphasized to be a genuine property. We then discuss the use of the average action and provide the analytical expressions that characterize the flow for various regulators.

Finally, we draw our conclusions in section V.

To summarise, the present paper is adressing the following issues :
\bei
\item How the flow behaves in the deep IR of the broken phase and in particular how the minimum selection is dynamically performed by the coarse graining.
\item How the perturbatively well-established phase boundary depends on the bare coupling $u$.
\item How the convexity of the effective potential is realized by the quantum fluctuations and the stability of the corresponding trajectories.
\item How universal are the trajectories in the deep IR.
\item How the addition of the regulator influences the previous results at LPA, as a preliminary step before the study of the wave function renormalization.
\item How the situation is changed in dimensions different than 2.
\eei

\section{Wegner-Houghton approach}
Let us start with a bare action defined at the UV cut-off $k$ :
\be
S_k[\Phi]=\int d^d x\,\le[\frac{1}{2}\d_\mu\Phi\d_\mu\Phi+V_k(\Phi)\ri]
\ee where the field configurations $\Phi(x)$ have non-zero  components only in the momentum-sphere of radius $k$.
We are interested in the effective theory for the low-frequency modes $\phi(x)$ which have momentum components only in the sphere of radius $k-\delta k$ once the fast modes $\vphi(x)$ carrying modes in the shell $[k-\delta k,k]$ have been integrated out. Splitting the original field into its low and fast components $\Phi(x)=\phi(x)+\vphi(x)$ and performing a one-loop expansion for the fast mode $\vphi$ yields :
\be\label{WH_action}
\frac{\delta S_{k}[\phi]}{\delta k}\delta k=S_k[\phi]-S_k[\phi+\vphi_0]-\frac{\hbar}{2}Tr\,Log'\le(\frac{\delta^2S}{\delta\vphi^2}\ri)+o(\hbar)
\ee where the $Tr'$ holds for a trace in the momentum space on a shell of $[k-\delta k,k]$ and is thus proportional to $\delta k$ and $\vphi_0$ is the saddle-point.
The blocked action described by the previous flow equation can easily be proven to be the effective action, see e.g. \cite{Alexandre:2009bq}.

The presence of the saddle point $\vphi_0$ in the flow equation opens the possibility to have tree-level renormalization \cite{Alexandre1999,Alexandre:1999sr}. This happens in particular in the IR of the broken phase for a $\phi^4$ scalar field theory in the internal region between the two minima where tree-level effects provide the convexity of the effective potential taking into account inhomogeneous field configurations. Let us mention that even if this effect helps us to obtain a convex potential for $k\to0$, this solution corresponds to a singularity of the flow equation with loop-evolution when the saddle-point is assumed to be trivial \cite{Alexandre1998,Pangon:2009pj}.
Such a tree-level has been expected to occur in a similar fashion in the broken phase of sine-Gordon model in d=2 \cite{Nagy:2006pq} and in $d\geq3$ \cite{Nandori:2003pk} in the concave regions using Fourier decomposition of the effective potential. We will show in the following that it is not the case for at least a part of the broken phase in $d=2$, the whole IR trajectories being beyond our numerical capacities. 

Finally, let us mention that the flow equation (\ref{WH_action}) for the action is so far a one-loop approximation. At this point, one can notice that the higher loops terms that have been neglected carry at least two momentum integrals, each of them being proportional to $\delta k$ so that in the exact differential limit $\delta k\to0$, the flow equation (\ref{WH_action}) turns out to be exact. As the functional determinant is of little use for practical purpose and is in general unknown for an arbitrary background field $\phi$, one defines the Wegner-Houghton equation to be the projection of the evolution equation for the action on a constant background.  As the blocking on sharp regions of the momentum space prevents us to define the next to leading order in the gradient expansion, see e. g. \cite{Bonanno1999,Polonyi:2001se}, the LPA is the only order of the gradient expansion that makes sense. The flow equation for the potential reads :
\be\label{WH_dimfull}
\frac{\d V_k(\phi)}{\d k}=-\frac{K_dk^{d-1}}{2}Log\le(1+\frac{V_k''(\phi)}{k^2}\ri)
\ee where $K_d$ is a geometrical factor given by :
\be
K_d=\frac{2\pi^{\frac{d}{2}}}{\Gamma(\frac{d}{2})(2\pi)^d}
\ee The equation (\ref{WH_dimfull}) holds as long as the curvature of the action at LPA is strictly positive in order to maintain the validity of the loop-expansion. The limit of applicability/reliability of this equation thus reads :
\be\label{WH_convexity}
1+\tv''(\phi)=0
\ee where $\tv''=V''/k^2$ and will be referred to as the convexity limit of the blocked action in the following.

\subsection{UV perturbative regime}
\subsubsection{Coleman Fixed point}\label{WH_pertu}

As we are looking for fixed points in the UV, let us define the flow equation for the dimensionless potential $\tv$ in $d=2$:
\be\label{WH_dimless}
k\frac{\d \tilde{V}}{\d k}+2\tilde{V}=-\frac{1}{4\pi} Log \le(1+\tv''\ri)
\ee where we can make the confusion between the dimensionless and dimensionfull quantities $\tp=\phi$ and $\tb=\beta$ due to a trivial scaling dimension.

The Coleman fixed point is defined to be a perturbative UV fixed point for a periodic potential when no wave-function renormalization is taken into account. Taking the bare potential :
\be
\tv_{\Lambda}=\tilde{u}_\Lambda cos(\beta \phi),\quad \tilde{u}_\Lambda \beta^2<<1,
\ee 
and linearizing the flow equation (\ref{WH_dimless}), one gets :
\be\label{flow_lin_WH}
4\pi k\frac{\d \tv}{\d k}+\tv ''+8\pi\tv=0
\ee where we neglected higher powers of $(\tv'')^n=O(\tilde{u}^n)$. This shows that the Coleman fixed point is characterized by the critical frequency :
\be\label{Coleman_WH}
\beta_c=\sqrt{8\pi}
\ee which reproduces the well-known result \cite{Coleman:1974bu}. 

Defining $\beta_r=\frac{\beta}{\beta_c}$, the UV behavior in the weak coupling regime is easily established :
\be\label{WH_coupling}
k\frac{\d \tilde{u}_k}{\d k}=2\tilde{u}_k\le[\frac{\beta^2}{\beta_c^2}-1\ri]=2\tilde{u}_k(\beta_r^2-1)
\ee
As a result, one expects that for $\beta_r>1$ the periodicity is unbroken due to vanishing of the dimensionless coupling with the flow while in the broken phase $\beta_r<1$ the dimensionless coupling increases with the flow i. e. it exhibits UV asymptotic freedom. Taking into account that the flow generates during the blocking higher frequency modes, it is straightforward to check that the critical frequency for the higher Fourier modes $n\beta$ is given by $\beta_r=1/n$ within this linearization \cite{Nagy:2006pq} so that when $\beta_r=0.5$, the Fourier mode of frequency $2\beta$ is marginal.

On the contrary, in both phases, the dimensionfull coupling vanishes :
\be
k\frac{\d u_k}{\d k}=2u_k\beta_r^2>0
\ee This is the signal that one expects a dimensionfull effective potential to be a constant in both phases being the only function periodic and convex. We conclude that the dimensionfull quantities are not good quantities to characterize the theory.

\subsubsection{Order parameter}
As the scale is decreased in the broken phase, one expects at some point that our approximation of small coupling regime breaks down. The upper bound of the dimensionless coupling is given by $\tilde{u}\to\frac{1}{\beta^2}$ which corresponds to the appearance of soft modes with a single harmonic ansatz. Indeed, in such a case, the flow equation (\ref{WH_dimless}) is given by :
\be\label{WH_soft}
k\frac{\d \tilde{V}}{\d k}+2\tilde{V}=-\frac{1}{4\pi} Log \le(1-\beta^2\tilde{u}cos(\beta\phi)\ri)
\ee where  soft modes have appeared in the logarithm every $\frac{2\pi}{\beta}$. In such a case, the linearization of the flow equation (\ref{flow_lin_WH}) is not valid anymore and one has to solve the full equation (\ref{WH_dimless}). 

The resolution for the Fourier expansion of the potential for the first 10 harmonics \cite{Nagy:2009pj} and references therein showed that the flow in the broken phase seems to drive the potential to be $\tv_k=-\frac{1}{2}\phi^2$ repeated periodically in the concave regions. This potential corresponds to the limit of  applicability of the loop-expansion used to derive the Wegner-Houghton equation (\ref{WH_dimfull}), making this scheme of renormalization questionable in such situation.
Inspired from our previous considerations, one concludes that one should monitor the dimensionless curvature of the action at the middle of the concave regions where the convexity limit (\ref{WH_convexity}) has the best odds to be violated :
\be\label{WH_top}
\tilde{S}''_k(\phi_n)=1+\tv''(\phi_n),\quad\phi_n=\frac{2n\pi}{\beta},\,n\in\N
\ee
At the perturbative level, this quantity has a $\beta$-function of opposite sign as the one of the sine-Gordon amplitude each side of the critical surface i. e. it is UV asymptotically free in the symmetric phase and positive in the broken one :
\be\label{curvature_beta}
k\d_k \tilde{S}''_k(\phi_n)=2\beta^2\tilde{u}_k(1-\beta_r^2)
\ee
In addition, it provides a kind of effective order parameter as :
\bei
\item in the symmetric phase, this goes to 1 as the flow suppresses the coupling.
\item in the broken phase, this a priori goes to $0$ if the system goes unstable in the IR.
\eei
This phenomenological order parameter has also the huge advantage to be easily computed in fRG, on the contrary to the topological susceptibility that has been demonstrated to be a natural order parameter for periodic theories \cite{Nandori:1999vi}.
If we are to make the connection with the SU(2) deconfinement transition, the curvature of the action has the opposite behavior than the one of the trace of the Polyakov loop : in the deconfined phase where the center symmetry is spontaneously broken, the trace of the Polyakov-loop is non-zero while it vanishes in the confined phase. 

\subsection{IR of the broken phase} 

\subsubsection{Weakly coupled bare theory}\label{WH_weakly_coupled}
\begin{figure}[h!]
\begin{center}
\includegraphics[angle=270,scale=0.5]{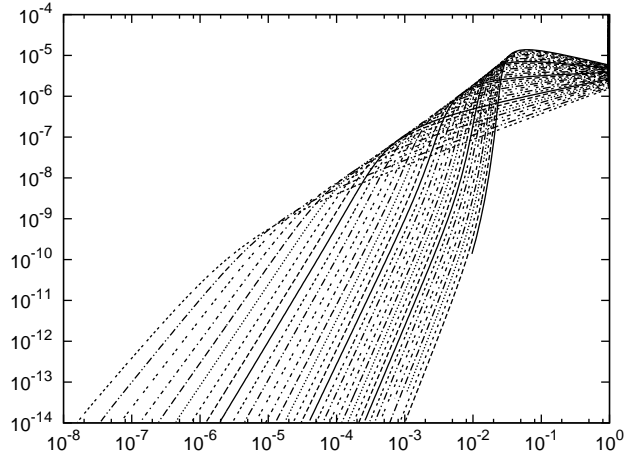}
\caption{The $\beta$-function of $\ts''(\phi_n)$ in units of $\Delta k$ as a function of the scale $k$ for $\beta^2\tilde{u}_\Lambda=0.01$ for increasing values of $\beta_r$ from $0.55$ (top in the UV, bottom in the IR)  to $0.90$ (bottom in the UV, top in the IR) by steps of $\Delta \beta_r =0.01$. One can identify a power-law decay of the $\beta$-function when reaching the neighborhood of the IR fixed points.}\label{WH_weak_beta}
\end{center}
\end{figure}

This study has been partially performed already in \cite{Pangon:2009wk} but we reproduce here refined results for completeness.
Let us first focus on the case where one starts with a perturbative initial condition $\beta^2\tu_\Lambda<<1$, i. e. the UV scaling is given by equation (\ref{WH_coupling}) holds. 
\begin{figure}[h!]
\begin{center}
\includegraphics[angle=270,scale=0.5]{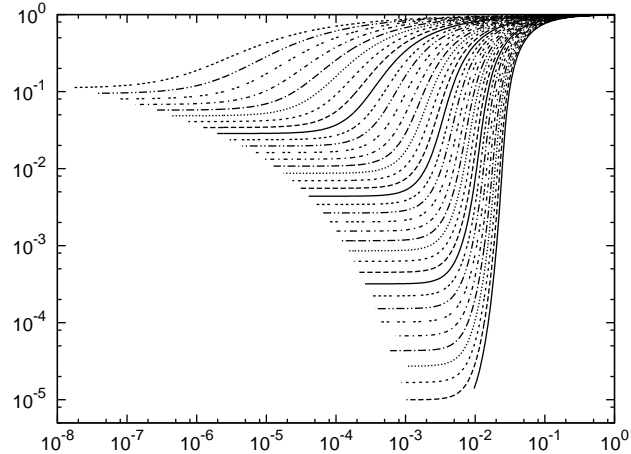}
\caption{The curvature at the middle of the concave regions (\ref{curvature_beta}) as a function of the scale $k$ for $\beta^2\tilde{u}_\Lambda=0.01$ for increasing values of $\beta_r$ from $0.55$ (bottom) to $0.90$ (top) by steps of $\Delta \beta_r =0.01$. The UV scaling is shortened for decreasing $\beta_r$, in relation with eq. (\ref{WH_coupling}).}\label{WH_weak}
\end{center}
\end{figure}
Starting with a "weak" bare amplitude $\beta^2u_\Lambda=0.01$, the IR of the whole broken phase exhibits a continuum of fixed points for $0.90>\beta_r\geq 0.55$, see fig. \ref{WH_weak}. This line of fixed-points ensures the stabilization of the loop expansion and thus the safety of the use of the Wegner-Houghton RG. They correspond to an highly non-trivial recombination of all the Fourier harmonics and thus can not be seen using a Fourier truncation of the effective potential. This point will be further illustrated in section \ref{WH_fourier}. A definite answer for lower frequencies than $\beta_r=0.55$ or larger than $\beta_r=0.90$ is unclear as one faces in the first case a very strong attraction of the convexity limit and a very low speed of the flow in the latter case. A study of the $\beta$-function for the curvature illustrates well these difficulties, see fig. \ref{WH_weak_beta}. We identify two different scaling regimes separated by a maximal speed of the flow, corresponding to the competition between the UV Coleman scaling and the IR fixed points.
The perturbative UV scaling is shortened as $\beta$ is decreased, due to a larger speed of the flow of $\tu$, see (\ref{WH_coupling}) and (\ref{curvature_beta}).
 
\begin{figure}[h!]
\begin{center}
\includegraphics[angle=270,scale=0.5]{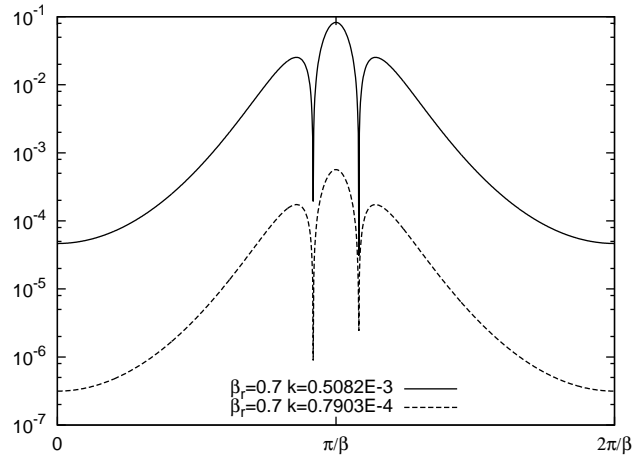}
\caption{The $\beta$-function of the absolute value of the curvature $|k\d_k \ts''(\phi)|$  as a function of $\phi$ for $\beta^2\tilde{u}_\Lambda=0.01$ for $0.7$ . The vertical lines correspond to a change of sign of the $\beta$-function : it is negative in the neiborhood of $\frac{\pi}{\beta}$, positive elsewhere.
For decreasing scales, the absolute value of the $\beta$-function decreases uniformely for all $\phi$.}\label{WH_flow_FP}
\end{center}
\end{figure}

\begin{figure}[h!]
\begin{center}
\includegraphics[angle=270,scale=0.5]{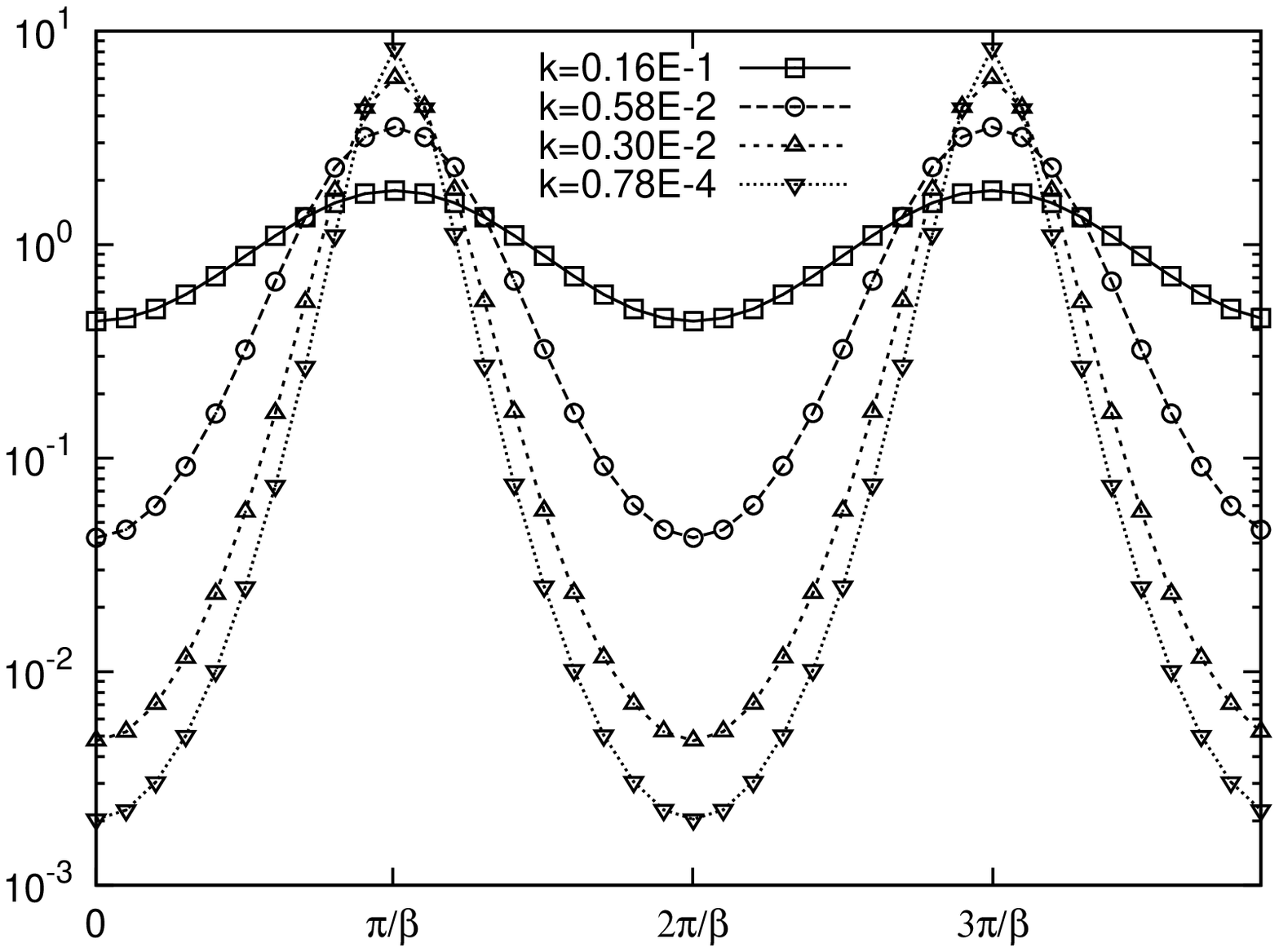}
\caption{The curvature of the action $\ts''(\phi)$ as a function of $\phi$ for four different scales for $\beta_r=0.7$ in d=2. The peak (resp. flat) regions correspond to the convex (resp. concave) regions for the potential, see fig. \ref{WH_potential_phi}. This shows that the peaks in the second derivative correspond to deep wells in the potential reflecting the progressive selection of a single vacuum, fluctuations around the latter being very massive.}\label{WH_curvature_phi}
\end{center}
\end{figure}

\begin{figure}[h!]
\begin{center}
\includegraphics[angle=270,scale=0.5]{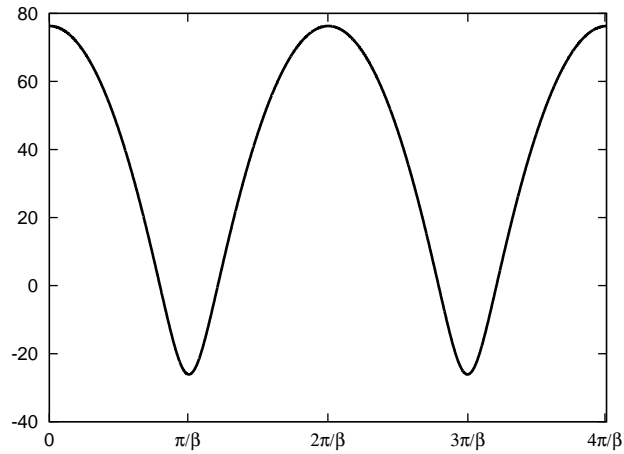}
\caption{The rescaled fixed point potential $\beta_c^2\tv$ with respect to $\phi$ for for $\beta_r=0.7$ in d=2. The concave regions where the curvature is flat are given by the upside-down parabola $-\frac{1}{2}0.998\beta_c^2\phi^2$, where $1-0.998=0.002$ is the minima of the curvature at $\phi=\phi_n$ of fig. \ref{WH_curvature_phi}.}\label{WH_potential_phi}
\end{center}
\end{figure}

For $\beta_r$ above $0.90$, the transition between the two regimes is shallow and the $\beta$-function is not enough suppressed relatively to the UV scaling to draw a clear picture. 
For $\beta_r$ below $0.55$, the large bump in the $\beta$-function signals the strong attraction of the singularity of (\ref{WH_convexity}) on the flow but the numerical errors prevent us to stabilize the IR scaling to establish undoubtedly -even if a strong suspicion is raised by the figure \ref{WH_weak}- the existence of an IR fixed point.
An important aspect here is that the fixed points exhibited here are not only fixed points of the curvature $\ts''_k(\phi)$ in $\phi=\phi_n$, but also fixed points for the whole range of $\phi\in \R$. Let us illustrate it in the special case $\beta_r=0.70$, see figure \ref{WH_flow_FP}. The change of sign of the $\beta$-function around $\frac{\pi}{\beta}$ is easily understood when monitoring directly the curvature, as shown on the figure \ref{WH_curvature_phi}. 
It displays the curvature as a function of $\phi$ for different scales. As it is well known, the spectrum of the broken phase of sine-Gordon is made of massive excitations, so that the dimensionless curvature in the convex regions is strongly enhanced as the scale is decreased below their typical mass scale. This point will be further discussed in section \ref{sect5}.
In the same time, the curvature vanishes in the concave regions reflecting the progressive decoupling between the different homotopy classes of the dual XY model. These two effects in convex and concave regions are stronger as $\beta_r$ is decreased.
Finally, the IR dimensionless effective potential is plotted on fig. \ref{WH_potential_phi}, where we see that the maxima of the curvature of fig. \ref{WH_curvature_phi} corresponds to deep minima of the potential, providing an effective mass term, while the minima of the curvature give us the upside-down parabola in the concave regions of the potential.

\subsubsection{Fourier expansion convergence}\label{WH_fourier}
Since the potential is periodic in the UV and since the equation (\ref{WH_dimless}) preserves periodicity, one could have expected a Fourier series to provide a sensible expansion for the potential and try to limit ourself in following the scale evolution of the couplings of the first Fourier components. Then, to ensure that we describe effectively the genuine solution, one could have checked the stability of such truncation when taking into account more harmonics. This strategy, followed so far in all the studies of sine-Gordon model is shown here to be misleading in the IR of the broken phase, in the same fashion as the polynomial expansion is in the broken phase of $\phi^4$-model \cite{Alexandre1998}.

For the reasons mentioned previously, one solves the flow equation for the second derivative of the potential. In principle, as it is expected to be periodic and continuous, Dirichlet theorem should ensure us at least a simple convergence of the Fourier serie towards the genuine solution when increasing the number of modes taken into account. In the symmetric phase, this convergence is expected to be very fast as the lowest frequency is suppressed by the flow and thus has little possibility to generate higher frequency modes which are themselves even more irrelevant. This is indeed what is found in \cite{Nagy:2006pq} and even a single harmonic ansatz for the potential could do the job.

On the contrary, the broken phase is more involved. First, the original lowest frequency is relevant and is enhanced by the flow, generating higher frequencies which may be themselves enhanced by the flow (if $\beta_r<0.5$) but are in any case less irrelevant as in the symmetric case. 
In addition, the higher frequency modes have an enhanced weight in the second derivative of the potential, their amplitude being the one of the potential times $p^2$ for the $p$-th mode.
\\But the largest source of error and the origin of the bad convergence of the Fourier expansion comes from a kind of Gibbs phenomenon in the concave regions of the potential. To illustrate this, let us define the Fourier expansion of our numerical solution :
\be\label{WH_dirichlet}
\tv''(\phi)=\sum_{p=1}^{n}c_p cos(p\beta\phi)
\ee where the $c_p$'s can be reconstructed from the usual identity :
\be\label{WH_fourier_der2}
c_p(k)=\frac{\beta}{\pi}\int_{-\frac{\pi}{\beta}}^{+\frac{\pi}{\beta}} \tv_k''(\phi)\,cos(p\beta\phi)d\phi
\ee
Let us illustrate for the special case $\beta_r=0.7$, $\beta^2u_\Lambda=0.01$. This initial condition corresponds to a trajectory entering the broken phase. Starting from the UV, the curvature weakly decreases and is then strongly attracted by the convexity limit (\ref{WH_convexity}) in the IR. For the second derivative of the potential, it amounts in the appearance in the IR of flat regions of constant value $\simeq-0.998$ where the potential is concave and peaks in the convex regions, see figures \ref{WH_curvature_phi}-\ref{WH_potential_phi}. 
\begin{figure}[h!]
\begin{center}
\includegraphics[angle=270,scale=0.5]{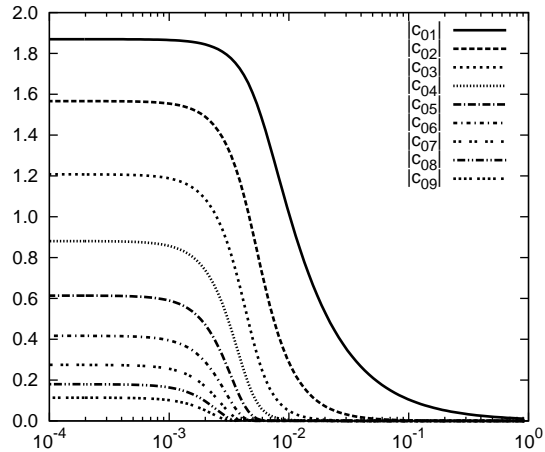}
\caption{The absolute value of the reconstructed Fourier coefficients of the second derivative of the potential defined by eq. (\ref{WH_fourier_der2}) as a function of the scale $k$ for $\beta_r=0.7$ and $u_{\Lambda}=0.01$. The even modes are negative and the odd ones positive.}\label{WH_fourier_beta0.7}
\end{center}
\end{figure}
Let us notice that this shape can be obtained only by the infinite resummation of all Fourier components as the higher order harmonics are IR enhanced. On figure \ref{WH_fourier_beta0.7}, one sees that when the original coupling becomes large enough, it triggers the build up of the higher modes, one after another. Since the sign of their contribution depends on the parity of the modes, the fixed point solution is a finely-tuned competition of all the modes.
This situation is comparable to the broken $\phi^4$ theory where we know one needs the exact resolution for the effective potential without any polynomial ansatz to recover the Maxwell-construction when projecting on the polynomial couplings \cite{Pangon:2009pj}.

\begin{figure}[h!]
\begin{center}
\includegraphics[angle=270,scale=0.5]{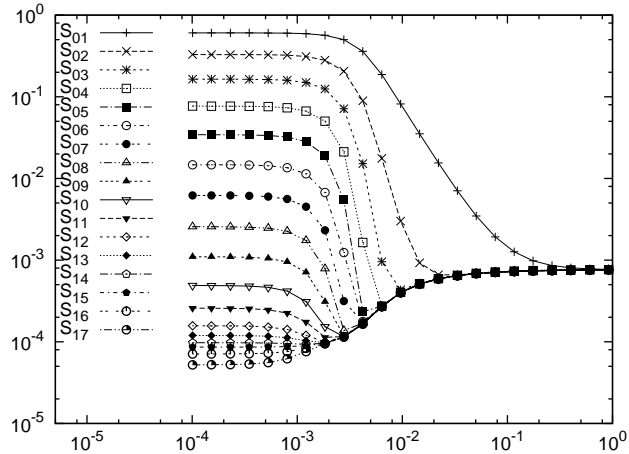}
\caption{The relative error of the Parseval sums $\le|\frac{S_{n}-I_k}{I_k}\ri|$ as a function of the scale $k$, see eqn. (\ref{WH_parseval_eq}). The first harmonic is a good description only in the UV perturbative regime, up to $k\simeq 0.1$ and represents less than 40\% of the genuine solution in the IR, when the fixed point is reached.  The relative error decreases monotonously as $n$ is increased.}\label{WH_parseval}
\end{center}
\end{figure}

 Even if the transition between the convex and concave regions has to be continuous, it is in practice rather sharp due to the logarithmic enhancement, cf (\ref{WH_dimless}). As a result, the Fourier serie tries to mimic a change in the second derivative that resembles a step-function, see fig. \ref{WH_curvature_phi} (note the log y-axis). The serie of the cosine-polynomial that are a set of continuous functions can not hope to converge uniformly towards an almost discontinuous function near the frontiers between convex and concave domains of the potential. Thus the partial sums of eq. (\ref{WH_dirichlet}) overshoot the second derivative of the potential and this overshoot does not die out \cite{Zygmund1955}. As a result, the truncation of the flow equation on the lowest frequency modes introduces spurious oscillations and is not able to stabilize before the convexity limit (\ref{WH_convexity}).
\begin{figure}[h!]
\begin{center}
\includegraphics[angle=270,scale=0.5]{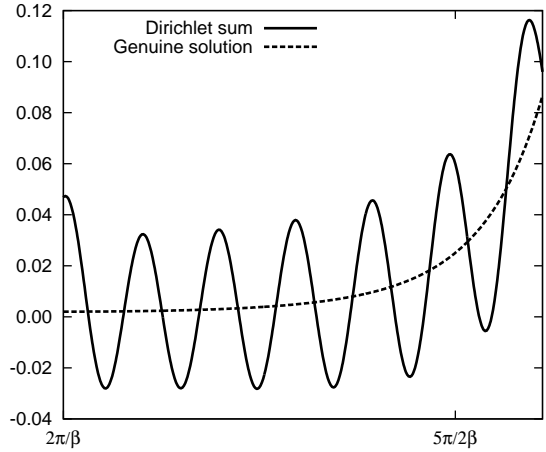}
\caption{The curvature of the action $\ts''$ as a function of $\phi$ for the scale for $k=0.79\cdot10^{-4}$ at which the fixed point solution has been reached in d=2 close to the edge of the flat region. The Fourier-Dirichlet sum stands for the partial sum of the Fourier serie up to order $n=17$, see eq. (\ref{WH_dirichlet}). The genuine solution is the solution coming from the resolution of the flow equation without Fourier expansion.}\label{WH_curvature_absolute_single}
\end{center}
\end{figure}
This is shown on figure \ref{WH_curvature_absolute_single} where the Fourier oscillations are shown to be violating the degeneracy criterion (\ref{WH_convexity}) while the genuine solution does not. Let us mention that this overshoot is  unseen using the Parseval identity :
\be\label{WH_parseval_eq}
I_k=\frac{\beta}{2\pi}\int_{-\frac{\pi}{\beta}}^{+\frac{\pi}{\beta}} |\tv_k''(\phi)|^2d\phi=\sum_{p=1}^{n\to\infty}|c_p(k) |^2=\lim_{n\to\infty}S_n
\ee
see figure \ref{WH_parseval}. The discrepancy is shown in the latter figure to die out as the integral of the genuine solution is not related to its continuity properties.

Due to this phenomenon, the Fourier expansion in the broken phase is of dangerous use, at least in the IR. From our point of view, this explains why the other studies  \cite{Nagy:2009pj} and references therein incorrectly predicted the appearance of the singularity and failed to see the fixed points line as the overshoot drove the flow in the unstable region for the flow equation.
The absence of the instability -the breakdown of loop-expansion- is of crucial importance in our opinion. If it was a genuine instability, it would mean first that the RG scheme used to find it was inapplicable in this case. But this would also signal that the related path integral sees equal contributions of  all field-configurations in the concave regions, a situation that can not be handled analytically nor numerically so far.
Having clarified this point, we present in the following the solutions of the flow equation without using any Fourier ansatz, see appendix \ref{app_num}.

\subsection{Universality in the broken phase}
We showed in the previous section that at least a part of the broken phase was having IR non-perturbative fixed points when the bare theory is chosen to be perturbative. 
A natural question is thus to determine in which extent the coupling $\tu_\Lambda$ has to be small and study the influence of its bare value on the existence of the IR fixed points and -if still exist- how the way they are reached changes. 

It is well known  when the wave-function renormalization is taken into account perturbatively  that for large enought couplings, the presence of a separatrix in the phase diagram places the system in a cross-over regime \cite{Kosterlitz:1973xp} and even simple ansatz in fRG reproduces it well, \cite{Nagy:2009pj}. 
\\On the other hand, if the initial coupling is really large, the perturbative linearization of the flow (\ref{flow_lin_WH}) breaks down as non-perturbative effects arise at the cut-off scale and no simple analytical method can locate the phase boundary. In such situation, an LPA study might teach us something valuable as we have in any case no argument to put forward to claim that we are in the cross-over regime that depends crucially on the wave-function renormalization. Starting with $\beta^2u_\Lambda\lesssim 1$ i.e. close to the break down of the loop-expansion in the UV, one expects the strong enhancement in the UV of the higher frequency modes in the same way as in the IR of the broken phase. As these modes have a relative weight larger in the second derivative of the potential -the $p$-th mode is $p^2$ enhanced- they can compete in the $\beta$-function of the curvature with the original mode. Since this happens already in the UV, one should expect non-trivial effects to appear and even possibly a change in the phase diagram. Let us show that it is not the case at LPA order.
As usual, we illustrate the results with the case $\beta_r=0.7$. 
\begin{figure}[h!]
\begin{center}
\includegraphics[angle=270,scale=0.4]{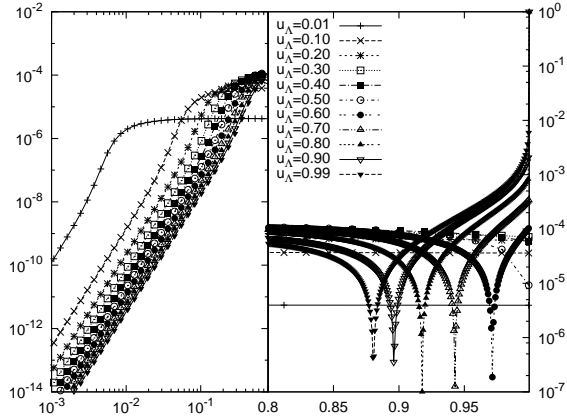}
\caption{The $\beta$-functions for the curvature $\ts''(\phi_n)$ in units of $\Delta k$ for $\beta_r=0.7$ for $\beta^2\tilde{u}_\Lambda\in[0.1:0.99]$.}\label{WH_strong_beta_u1_impact}
\end{center}
\end{figure}
On figure \ref{WH_strong_beta_u1_impact}, one sees that in the UV the flow of the curvature tends to restore the symmetry for large enough bare coupling, at odd of the perturbative relation (\ref{curvature_beta}), then the curvature saturates and finally reaches an IR fixed point. For smaller values of the bare coupling, the $\beta$-function remains positive all along the trajectory in agreement with the perturbative expectations, eq. (\ref{curvature_beta}). The power-law scaling found in the IR is a universal quantity, characterized by an exponent independent of $\tilde{u}_\Lambda$ (but dependent on $\beta_r$).

\begin{figure}[h!]
\begin{center}
\includegraphics[angle=270,scale=0.4]{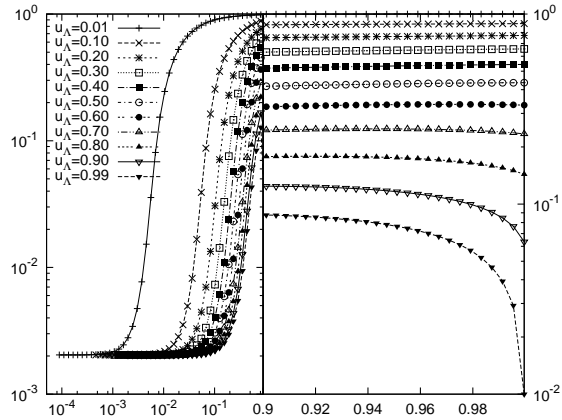}
\caption{The curvature $\ts''(\phi_n)$ with respect to the scale $k$ for $\beta_r=0.7$ for $\beta^2\tilde{u}_\Lambda\in[0.1:0.99]$.}\label{WH_strong_u1_impact}
\end{center}
\end{figure}

\begin{figure}[h!]
\begin{center}
\includegraphics[angle=270,scale=0.4]{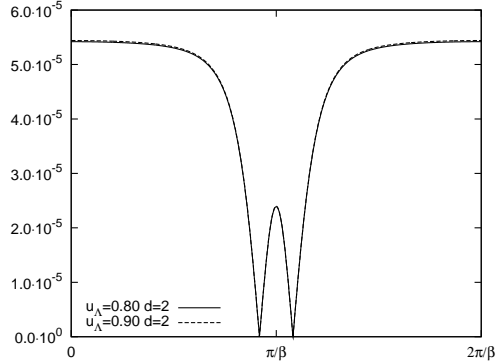}
\caption{The relative error on the deep IR curvature $\ts''(\phi)$ with respect to $\phi$ for $\beta_r=0.7$ for $\beta^2\tilde{u}_\Lambda=0.8$ and $0.9$. The reference is chosen to be the fixed point curvature computed for $\beta^2\tilde{u}_\Lambda=0.01$}\label{WH_curvature_relative_strong}
\end{center}
\end{figure}

\begin{figure}[h!]
\begin{center}
\includegraphics[angle=270,scale=0.4]{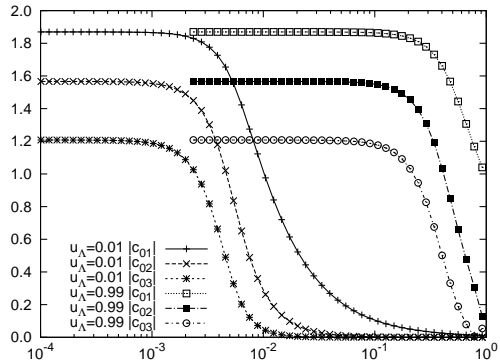}
\caption{The Fourier coefficients $c_p$ of definition (\ref{WH_fourier_der2}) for $\beta_r=0.7$ for $\beta^2\tilde{u}_\Lambda=0.01$ and $0.99$. The fixed point structure is the same in both cases, but the flow history is different. In the strong bare coupling case, the higher frequency modes are generated already in the UV.}\label{WH_strong_fourier}
\end{center}
\end{figure}

\begin{figure}[h!]
\begin{center}
\includegraphics[angle=270,scale=0.4]{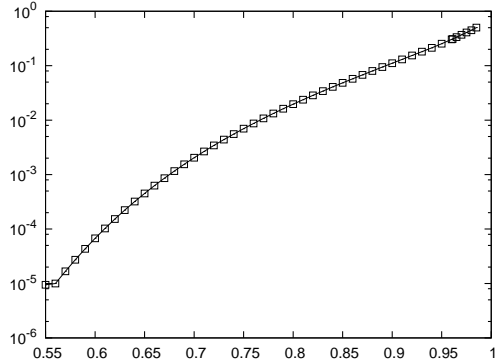}
\caption{The fixed point structure of the broken phase i. e. the IR value of the curvature as a function of $\beta_r$. The values for $\beta_r>0.90$ come from the conjecture that strongly coupled bare theory also converge to the same IR fixed as the weakly coupled one.}\label{WH_phase_summary}
\end{center}
\end{figure}

More interestingly, the figure \ref{WH_strong_u1_impact} shows that not only the approach to the fixed points is universal, but also the value of the curvature itself is independent on the bare coupling. Let us mention that this generalized universality spreads for the whole range of $\phi$, as exhibited on figure \ref{WH_curvature_relative_strong}, where the relative error on the IR curvature for different values of $\tu_\Lambda$ is shown to be very small. The specific shape of the plot is driven by the shape of the not-exactly zero $\beta$-function of the curvature at finite $k$.
A study of the Fourier couplings demonstrates immediately that the UV enhancement of the curvature for strong enough bare coupling is due to the dominance of the higher frequency modes that are strongly generated in the UV, see figure \ref{WH_strong_fourier}. 
The property of generalized universality -the presence of the same IR fixed points at strong and weak coupling- has been checked to occur for any $\beta_r>0.55$. For lower $\beta_r$, the problem of facing the too strong attraction of the instability (\ref{WH_convexity}) is obviously still present. Noticing that the final effect of choosing a strong bare coupling amounts only in reaching faster the same IR fixed point, one can reasonably conjecture that for $1>\beta_r>0.90$ the fixed point structure is preserved and allow us to find the fixed point value for the curvature for $0.985>\beta_r>0.90$ using a strongly coupled bare theory. As a result, one can summarize the set of fixed points of the broken phase in figure \ref{WH_phase_summary}. A clear answer about the sectors $\beta_r<0.55$ and $\beta_r>0.985$ is beyond our numerical capacities.

\section{Average action}
As one knows that the effect of the renormalization of the temperature of the dual XY model is strong, one needs to capture the main features of the sine-Gordon model to take into account the wave-function renormalization that amounts here in the effective change with the scale of the $\beta$-frequency. Unfortunately, the Wegner-Houghton RG which belongs to the class of RG with sharp regulator does not make sense at the next to LPA order. To overcome this difficulty, one needs another RG scheme with a smooth regularization. The average action method \cite{Wetterich:1992yh}, being an "exact" scheme i.e. not based on any kind of approximation is the best candidate. We study in this section the price to pay for the use of the regulator at LPA where we can compare it to Wegner-Houghton as a preliminary work before addressing wave-function renormalization

Let us start with the bare action $S_\Lambda[\phi]$. The goal is to build a functional that interpolates between the bare action and the full regular effective action by a progressive turning on of the fluctuations with the help of a regulator playing the role of a momentum dependent mass term $M_k[\phi]=\frac{1}{2}\int \phi R_k \phi$. The partition function of this modified theory now reads :
\be\label{AA_Z}
Z_k=\int \D\phi\,e^{-\frac{S_\Lambda+M_k+J\phi}{\hbar}}
\ee
The generating functional of the connected n-points functions is thus :
\be
W_k=-\hbar Log Z_k
\ee
Let us define the so-called average action to be :
\be\label{AA_legendre}
\Gamma_k=min_{J}\le\{W_k-J\phic\ri\}-M_k[\phic]
\ee
This shows that $\Gamma+M$ is a convex quantity,  being the result of a Legendre transform. One can then derive the flow equation for the average action by formal functional  manipulations \cite{Wetterich:1992yh} where only the regularity of the Legendre transform (\ref{AA_legendre}) is needed. This assumption amounts in having $\Gamma_k^{(2)}+R_k>0$ for any scale $k$ during the flow. The resulting flow equation is then given by :
\be\label{AA_flow}
\frac{\d \Gamma_k}{\d k}=\frac{\hbar}{2} Tr\le(\frac{\d_k R_k}{R_k+\Gamma_k^{(2)}}\ri)
\ee

Let us now turn our attention to the sine-Gordon model particular case, where $S_\Lambda[\phi]$ is a periodic functional in the field $\phi$.  
The n-points functions computed from $W_k$ or $Z_k$ a priori correspond to a theory where periodicity is explicitly broken by the addition of $M_k$. On the other hand, one can gain some insights on these expressions using a one-loop evaluation of the partition function $Z_k$ of eq. (\ref{AA_Z}) which yields :
\be\label{AA_loop}
\Gamma_k=S_\Lambda[\phic]+\frac{\hbar}{2}Tr\,Log\le(S_\Lambda''[\phic]+R_k\ri)+o(\hbar)
\ee where the regulator contribution at tree-level has cancelled out.
This loop expansion is certainly valid in the UV as the regulator dominates and behaves as a large mass term \cite{Wetterich:1992yh}. This shows that the initial condition for (\ref{AA_flow}) can be safely chosen to be periodic. Then the flow equation preserves the periodicity as  the regulator is involved only in $\Gamma^{(2)}+R_k$ i. e. it adds an effective mass to the fluctuations. This situation, comparable to the massive sine-Gordon model \cite{Nagy:2006ue}, drives us to conclude that no further non-periodicity than the one from the regulator (that will eventually vanish at $k\to0$) will appear. In particular, this allows us to study the second derivative of the average action with respect to $\phi$ as a periodic function of $\phi$.
\subsection{UV perturbative regime}
The Local Potential Approximation of the equation (\ref{AA_flow}) reads in $d$ dimensions :
\be\label{AA_LPA}
\frac{\d V_k}{\d k}=\frac{1}{2}\int d^dp \frac{\d_k R}{R+p^2+V''}
\ee
It is convenient to express the regulator as :
\be
R_k=p^2 r(\frac{p^2}{k^2}),\quad k\d_kR=-2\frac{p^4}{k^2} r'(\frac{p^2}{k^2})
\ee where the function $r$ is constrained by the regulator prescriptions \cite{Wetterich:1992yh} :
\be\label{AA_constraints}
\lim_{y\to 0 }r(y) =\infty,\quad\lim_{y\to \infty }r(y) =0
\ee
This allows to rewrite (\ref{AA_LPA}) in $d=2$ using the dimensionless variable $y=\frac{p^2}{k^2}$ :
\be\label{AA_flow_dimless}
\d_t \tv +2\tv =-\frac{K_2}{2}\int_0^{\infty}dy\frac{y^{2}r'(y)}{yr(y)+y+\tv''}
\ee
Following the same strategy as in the Wegner-Houghton case, we can linearize  the flow equation in the UV :
\beq\label{flow_lin_AA}
\d_t \tv +2\tv 
&=&-\frac{1}{4\pi}\int_0^{\infty}dy\frac{yr'(y)}{r(y)+1}\nn
&+&\tv''\frac{1}{4\pi}\int_0^{\infty}dy\frac{r'(y)}{(r(y)+1)^2}
\eeq 
where the constant field independent contribution of the right hand side of  the first line is a left-over of the regulator and can be discarded here as a shift of the vacuum energy.
From the linearized flow in the UV, eq (\ref{flow_lin_AA}) one can check that the Coleman frequency is the usual one:
\be\label{Coleman_AA}
\beta_c^2=\le(-\frac{1}{8\pi}\int_0^{\infty}dy\frac{r'(y)}{(r(y)+1)^2}\ri)^{-1}=8\pi
\ee
This result has been obtained independently of the choice of the regulator and its parameterization, only using the general properties eq. (\ref{AA_constraints}).
The considerations from the WH case generalize immediately and  the harmonic $cos(\beta\phi)$ is found to be irrelevant for $\beta>\beta_c$, and relevant for $\beta<\beta_c$.

Let us now introduce the regulators that will be used in following. 
The regulator (sometimes called Litim's \cite{Litim:2000ci}) is a very popular choice as the loop-integral (\ref{AA_flow_dimless}) can then be computed analytically :
\beq\label{Coleman_Litim}
r_l(y)&=&\le(\frac{1}{y}-1\ri)\Theta(1-y)\nn
k\d_k\tv +2\tv&=&\frac{2}{\beta_c^2}\frac{1}{1+\tv''}
\eeq
If we are to prepare the inclusion of wave-function renormalization, we will also need a smooth regulator, the Litim regulator having trouble with gradient expansion \cite{PhysRevD.67.065004}. We pick for simplicity the power-law defined by :
\beq\label{regulators}
r_p(y)&=&a_py^{-b_p},\quad (a_p,b_p)\in \R^{+*}\times [1:\infty]
\eeq  where the loop-integal (\ref{AA_flow_dimless}) can be analytically performed, see \cite{Nandori:2009ad} and appendix \ref{app_power}. 
This regulator for the particular choice $b_p=1$ i. e.  $R_k=k^2$  is actually a functional Callan-Symanzik RG scheme \cite{Litim:1998nf,Alexandre:2000eg,Alexandre:2001wj} and the corresponding flow equation can be recasted into the Wegner-Houghton equation (\ref{WH_dimless}), see \cite{Nandori:2009ad} and appendix \ref{app_internal}. This alternative derivation for Wegner-Houghton from a scheme using a regularization from massive fluctuations provides us even more confidence that average action is able of dealing with periodic theories.

To summarize, we showed that even if the average action scheme introduces a mass term to regulate the fluctuations, this insertion is actually soft as this does not break explicitly the periodicity of the initial condition and the massive fluctuations are periodic. 
The phase boundary established in the UV was shown to be preserved by the addition of any regulator and we can hope that the IR is mostly untouched due to the vanishing of the regulator.

\subsection{IR of the broken phase}
\subsubsection{Weakly coupled bare theory}
\begin{figure}[h!]
\begin{center}
\includegraphics[angle=270,scale=0.5]{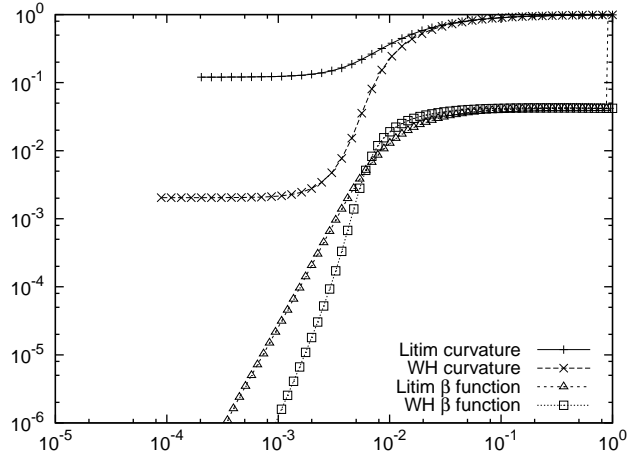}
\caption{The curvature and its $\beta$-function in units of $10^{-4}\Delta k$ with respect to the scale $k$ for $\beta^2\tilde{u}_\Lambda=0.01$ and $\beta_r=0.70$}\label{Litim_WH_0.7}
\end{center}
\end{figure}

Let us redo the $d=2$ study performed in section \ref{WH_weakly_coupled} for Wegner-Houghton RG with Litim's regulator for simplicity. We will see after that the qualitative and most of the quantitative results will spread to other regulators. Starting with a weakly coupled bare theory $\beta^2u_\Lambda=0.01$, we scan again the phase diagram for various $\beta_r$. The equivalent quantity of the dimensionless curvature for Litim regulator is :
\be
\ts''(\phi_n)=1+\tv''(\phi_n)
\ee
 which tests this time the possible singularity of the Legendre transform (\ref{AA_legendre}). This will be further discussed in the next section.
\begin{figure}[h!]
\begin{center}
\includegraphics[angle=270,scale=0.5]{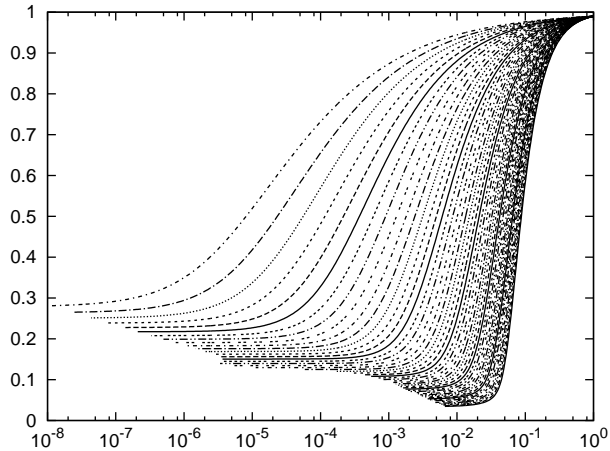}
\caption{The curvature $\ts''(\phi_n)$ with respect to the scale $k$ for $\beta^2\tilde{u}_\Lambda=0.01$ for increasing values of $\beta_r$ from $0.40$ (bottom) to $0.90$ (top) by steps of $\Delta \beta_r =0.01$.}\label{Litim_weak}
\end{center}
\end{figure}
The figure \ref{Litim_weak} exhibits the general tendencies of the flow for various $\beta_r$. The first obvious remark is that the curvature still exhibits a continuum of fixed points. This feature is well preserved by the addition of the regulator as it could have been expected, being a phenomenon occurring in the IR where the regulator vanishes. The second remark is that the fixed point value of the curvature is larger as in Wegner-Houghton RG (note the linear scale for the $y$-axis). The reason is that the regulator which behaves as a mass term suppresses a part of the fluctuations that were shown to decrease the curvature in Wegner-Houghton picture. Finally, one can notice that we are able to stabilize a fixed point for lower $\beta_r$ values, namely around $\beta_r=0.4$ (one can stabilize few lower trajectories not displayed for clarity). Realizing that the curvature is still "large", we discover here that the convexity limit $1+\tv''=0$ is not the limiting phenomenon as it was the case in Wegner-Houghton picture. This will be shown to be related to the non-differentiability limit of the running potential in the convex regions when using Litim regulator.
\begin{figure}[h!]
\begin{center}
\includegraphics[angle=270,scale=0.5]{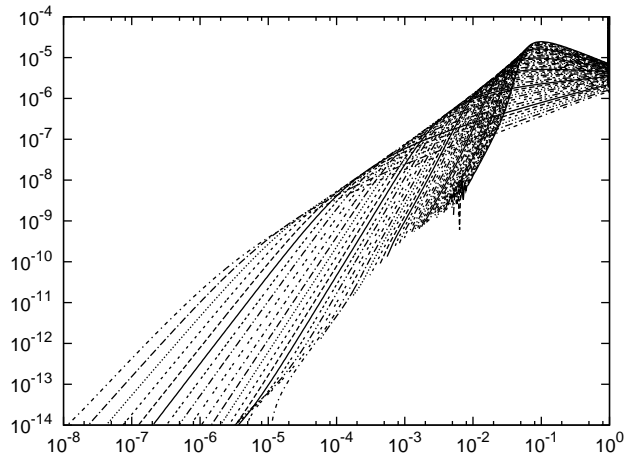}
\caption{The $\beta$-function of the curvature $\ts''(\phi_n)$ in units of $\Delta k$ with respect to the scale $k$ for $\beta^2\tilde{u}_\Lambda=0.01$ for increasing values of $\beta_r$ from $0.40$ (top in the UV, bottom in the IR) to $0.90$ (top in the IR, bottom in the UV) by steps of $\Delta \beta_r =0.01$.}\label{Litim_weak_beta}
\end{center}
\end{figure}
These considerations are further supported by the $\beta$-function of the curvature, see figure \ref{Litim_weak_beta}. The general behavior of the $\beta$-function is the same as the one obtained in Wegner-Houghton namely the existence of a maximal speed deep-enough in the broken phase corresponding to the cross-over between Coleman and IR fixed points. The IR scaling regime when entering the fixed point neighborhood is also given by a power-law decay. Interestingly, the power is different to the one obtained in Wegner-Houghton case. The figure \ref{Litim_WH_0.7} shows that the UV scaling regime is the same, being given by perturbation theory in both cases. Around $k=10^{-2}$ the flow is entering the cross-over regime and Litim and Wegner-Houghton flows start to diverge. The latter is strongly attracted by the convexity limit while the former still feels the influence of the regulator suppressing the fluctuations. As a result, the trajectories end at different fixed point values and the fact that the power of the decay towards the fixed point are slightly different shows that the regulator impact at higher scales is not totally washed out.
\subsubsection{Fourier expansion convergence and convexity issue}
We saw in the Wegner-Houghton case that the Fourier serie had a bad convergence in the IR of the broken phase due to the appearance of large flat regions of the second derivative of the potential corresponding to a dynamical Maxwell-cut in the concave regions. The loop-expansion has been shown to be stabilized before the occurrence of the instability for a part of the phase diagram $\beta_r>0.5$.
Using average action, the IR of a scalar broken phase in the concave region might also face an instability. This instability is not related to the degeneracy of the average action as the derivation of the flow equation does not rely on a loop-expansion but comes from the Legendre transformation (\ref{AA_legendre}) which needs $\Gamma^{(2)}_k+R_k$ to be invertible i.e. $\Gamma^{(2)}_k+R_k>0$. At LPA, the convexity limit is thus given by \cite{Tetradis:1992qt} :
\be\label{AA_convexity}
\overline{k}^2+V''=0
\ee where $\overline{k}^2=min_p\le\{p^2+R_k(p)\ri\}$. This limit has been found to be strongly IR  attractive \cite{Tetradis:1992qt} in the internal region of a broken $\phi^4$ model in $d=4$ giving rise to inhomogeneous tree-level contributions below the scale of appearance of this would-be instability. In such a case, the IR potential in the concave regions is given by the upside down parabola $V_k=-\frac{\overline{k}^2}{2}\phi^2$ and thus depends explicitly on the regulator. This raises the question of the possibility of removing the regulator for $k\to0$ in the internal region. Intuitively, one realizes this problem noticing that the regulator is a dimensionfull quantity so that it needs the appearance of another mass scale to decouple and then vanish. In the internal region, there is no other physical scale so that this decoupling can not occur at any finite scale $k$. The regulator being the only available scale in the non-convex region drives the flow and is itself considered as the typical momentum scale of the inhomogeneous configurations dominating the IR of the broken phase. Another way to say this is to notice that while the dimensionfull potential goes to 0 in the internal region -reproducing Maxwell construction-, the dimensionless one reaches a fixed point $\tv=-\frac{\tilde{k}^2}{2}\phi^2$ corresponding at the convexity limit where $\overline{k}^2=k^2\tilde{k}^2$. 
From these considerations, it is now obvious why the average action is able to provide the jump in the susceptibility at the border of the concave region : in the internal region the flow continues to stick to the regulator while on the contrary, the regulator has decoupled due to the presence of a mass term near the minima (the Higgs mass). The picture is even stronger in dimensionless quantities, when the second derivative reaches the fixed point $-\tilde{k}^2$ in the internal region and is quadratically enhanced in the external one ($\simeq\frac{m_H^2}{k^2}$).

\begin{figure}[h!]
\begin{center}
\includegraphics[angle=270,scale=0.5]{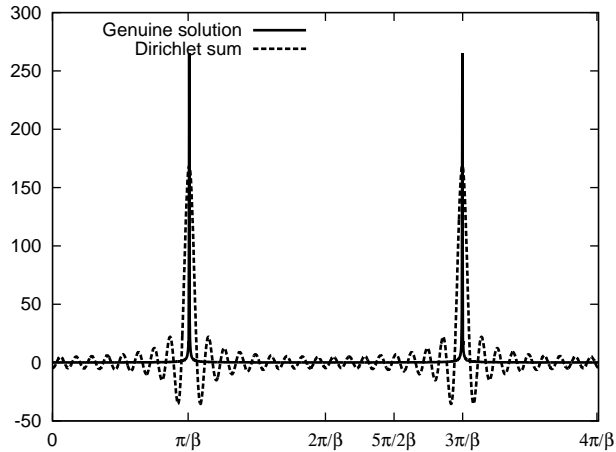}
\caption{The curvature of the action $\tilde{k}^2+\tv''$ with respect to $\phi$ for the scale for $k=0.25\cdot10^{-3}$ at which the fixed point solution has been reached. The Fourier-Dirichlet sum stands for the partial sum of the Fourier serie up to order $n=15$, cf. eq. (\ref{WH_dirichlet}). The genuine solution is the solution coming from the resolution of the flow equation without Fourier expansion.}\label{Litim_curvature_absolute_single}
\end{center}
\end{figure}

\begin{figure}[h!]
\begin{center}
\includegraphics[angle=270,scale=0.5]{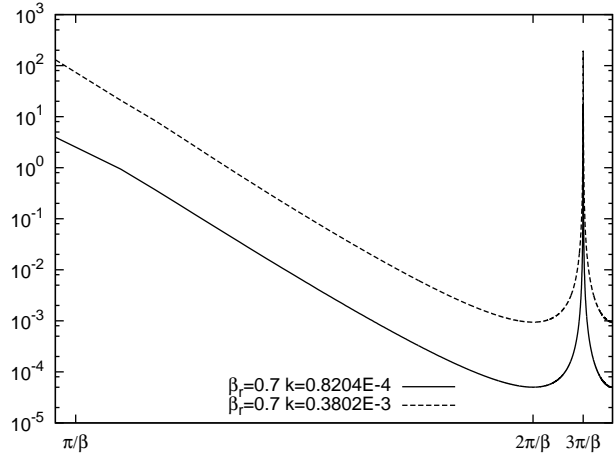}
\caption{The $\beta$-function of the absolute value of the curvature $|k\d_k \ts''(\phi)|$ with respect to $\phi$ at the end of the flow for $\beta^2\tilde{u}_\Lambda=0.01$ for $\beta_r=0.7$ for two different scales. A log-scale is used for the $x$-axis and the value $\phi=\frac{\pi}{\beta}$ is meant to be reached only asymptotically. The change of sign of the $\beta$-function in $\frac{\pi}{\beta}$ and $\frac{3\pi}{\beta}$ is hidden here for clarity as it is reduced to a vertical line.}\label{Litim_curvature_fixed_point}
\end{center}
\end{figure}

Let us now turn our attention to the sine-Gordon model. From similar considerations than the ones used in WH case, the well suited order parameter is : 
\be\label{AA_curv}
\ts''(\phi_n)=\tilde{k}^2+\tv''(\phi_n)
\ee where $\phi_n$ in defined by eq. (\ref{WH_top}).
Due to this jump in the susceptibility, the Gibbs phenomenon invoked in WH case is even more predominant being typically e.g. a power-law enhancement using Litim regulator (cf. eq. (\ref{Coleman_Litim})) to be compared the the logarithmic one in WH (cf. eq. (\ref{WH_dimless})). So that we expect the Fourier serie convergence to be even worse in the average action case. Let us illustrate this for $\beta_r=0.7$ in $d=2$ using Litim regulator for which $\tilde{k}=1$.
\begin{figure}[h!]
\begin{center}
\includegraphics[angle=270,scale=0.5]{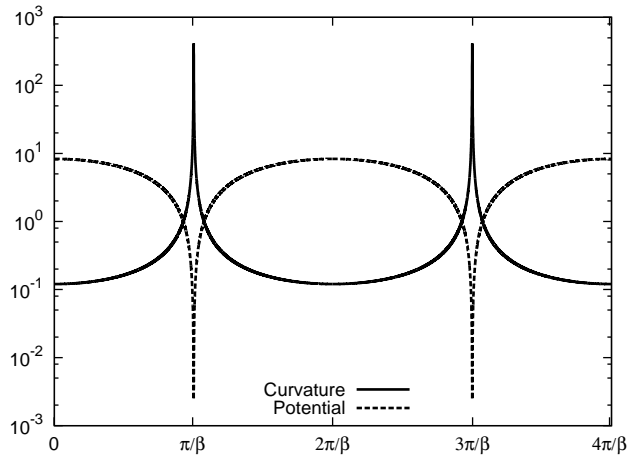}
\caption{The curvature $\ts''(\phi)$ with respect to $\phi$ for $\beta_r=0.7$ in d=2 (continuous line) for $k=10^{-4}$. The peak (resp. flat) regions correspond to the convex (resp. concave) regions for the potential. The dashed line corresponds to the rescaled potential $\beta_c^2\tv$ at the same scale and shows that the peaks in the second derivative correspond to deep wells in the potential reflecting the huge selection of a single vacuum. The concave regions where the curvature is flat are given by the upside-down parabola $-\frac{1}{2}0.88\beta_c^2\phi^2$, in relation with the fixed point value of fig. \ref{Litim_WH_0.7}.}\label{Litim_curvature_phi}
\end{center}
\end{figure}
In Figure \ref{Litim_curvature_phi}, one sees that in the IR, the wells in the potential selecting one minimum among the others is much stronger than in WH (note the logarithmic scale). 
As discussed previously, this is given by the jump of the susceptibility provided by the regulator. As a result, the comparison between genuine result and the Dirichlet partial sum provided by figure \ref{Litim_curvature_absolute_single} shows a much larger discrepancy compared to WH case. 
Obviously, the deeper we are in the broken phase, the stronger is the selection of the minima. This consideration shows that the limiting $\beta_r$ where we are able to find a fixed point is not given by the convexity limit but rather by the limit where in the IR the selection is so strong that it amounts in a numerically non-differentiable interaction. To illustrate this point, let us study the $\beta$-function for the whole curvature of figure \ref{Litim_curvature_fixed_point}. While the flow is clearly suppressed in the whole concave regions as the figure \ref{Litim_weak_beta} was suggesting it, the $\beta$-function is still strong in the convex regions when the flow is stopped by the non-differentiability of the interaction. On the other hand, we can notice that the $\beta$-function is decreasing fastly in the IR so that there is good hope that it will also reach a maximal value. We will see in the next section that the fact that this small scale dependence at the end of our flow has no impact on our conclusions.

\subsection{Universality in the broken phase}
Having found that the average action allows us to find the same set of fixed points as in the Wegner-Houghton case, it is tempting to test if the universality found previously still holds. This is actually far from obvious as picking a strongly coupled bare theory changes its relative weight to the regulator in the UV. We will demonstrate here that this difficulty does not occur.

\begin{figure}[h!]
\begin{center}
\includegraphics[angle=270,scale=0.4]{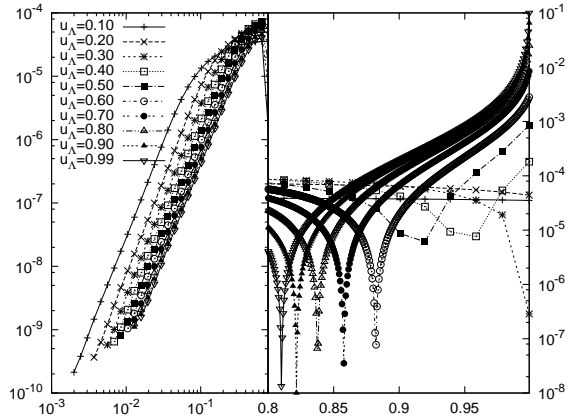}
\caption{The $\beta$-functions for the curvature $\ts''(\phi_n)$ in units of $\Delta k$ as a function of the scale $k$ for $\beta_r=0.7$ for $\beta^2\tilde{u}_\Lambda\in[0.1:0.99]$.}\label{Litim_strong_beta_u1_impact_beta}
\end{center}
\end{figure}

\begin{figure}[h!]
\begin{center}
\includegraphics[angle=270,scale=0.4]{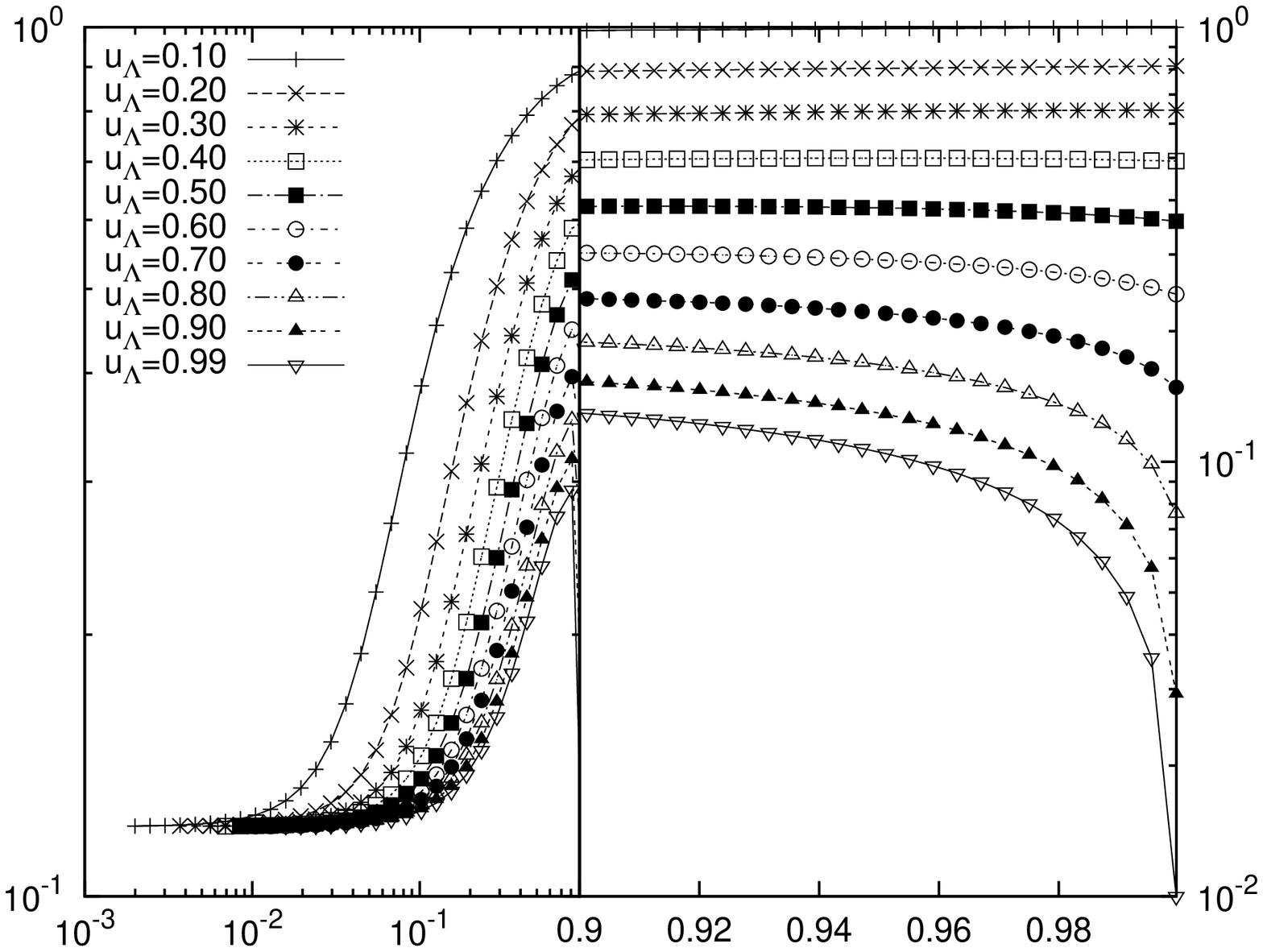}
\caption{The curvature $\ts''(\phi_n)$ as a function of the scale $k$ for $\beta_r=0.7$ for $\beta^2\tilde{u}_\Lambda\in[0.1:0.99]$.}\label{Litim_strong_beta_u1_impact}
\end{center}
\end{figure}

\begin{figure}[h!]
\begin{center}
\includegraphics[angle=270,scale=0.5]{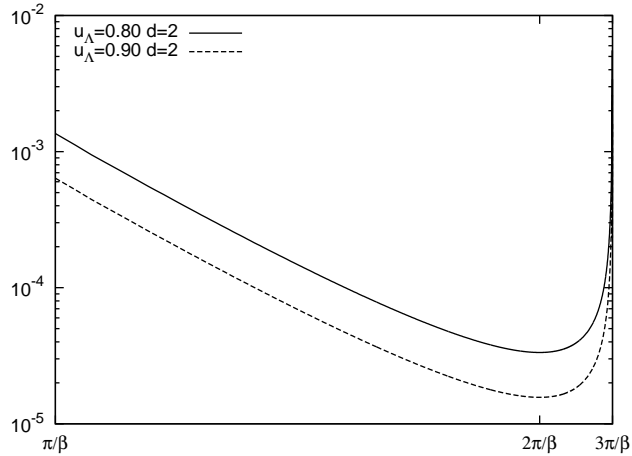}
\caption{The error on the fixed point curvature  for $\beta_r=0.7$ for $\beta^2\tilde{u}_\Lambda=0.80$ and $0.90$ in $d=2$ relatively to the curvature computed for $\beta^2\tu_\Lambda=0.01$ at the end of the flow.  A log-scale is used for the $x$-axis and the value $\phi=\frac{\pi}{\beta}$ is meant to be reached only asymptotically.}\label{Litim_curvature_relative_coupling}
\end{center}
\end{figure}

The figure \ref{Litim_strong_beta_u1_impact_beta} shows the $\beta$-function of the curvature of the theories for $\beta_r=0.7$ and various $u_\Lambda$. For couplings large enough, the $\beta$-function of the curvature also changes sign along the trajectory and exhibits a common power-law decay towards the fixed point independently of $\tu_\Lambda$.
The figure  \ref{Litim_strong_beta_u1_impact} shows that the fixed points  values are also independent of $\tu_\Lambda$, in the same fashion as in WH case.
The figure \ref{Litim_curvature_relative_coupling} allows us to check that the whole curvature, not only at the middle of the concave regions, is the same for different large bare couplings.

Obviously, all these results can be generalized to other $\beta_r$ values and as a strong bare coupling amounts only in a faster convergence to the fixed point, one can establish its presence for $0.90<\beta_r<0.98$.

\begin{figure}[h!]
\begin{center}
\includegraphics[angle=270,scale=0.5]{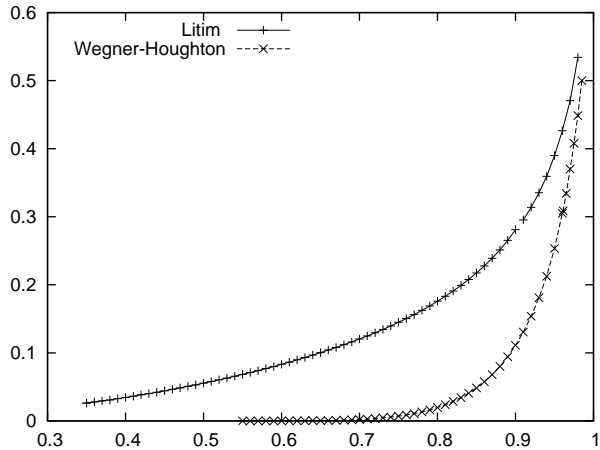}
\caption{The fixed point curvature using average action with Litim regulator in function of $\beta_r$. The Wegner-Houghton results are displayed for comparison.}\label{Litim_phase_summary}
\end{center}
\end{figure}

Finally, one can summarize the fixed point structure of the broken phase in figure \ref{Litim_phase_summary} where the difference with Wegner-Houghton is clear. Let us notice that the fixed point curvature as a function of $\beta_r$ has a clearly different behavior in the neighborhood of $\beta_r\simeq 1$.

\subsection{Influence of the regulator}
One of the final goal of this study being to prepare the situation for the inclusion of the wave-function renormalization, we have to be sure that the results established previously still hold for a smooth regulator and are not an artifact of the regulator choice. In this sense, this section can be seen as the first steps of an optimization study.

As expected the flow equation to solve is much more time consuming being an integro-partial differential equation as the momentum integral (\ref{AA_LPA},\ref{AA_flow_dimless}) can not be performed analytically in general. Let us present first the results for a power-law regulator given by (\ref{regulators}) for $a_p=1$ and $b_p=2$, see \cite{Nandori:2009ad} and appendix \ref{app_power}. In such a case, one has to use $\tilde{k}^2=2$ in the definition of $\ts''(\phi_n)$, see (\ref{AA_curv}).

\begin{figure}[h!]
\begin{center}
\includegraphics[angle=270,scale=0.5]{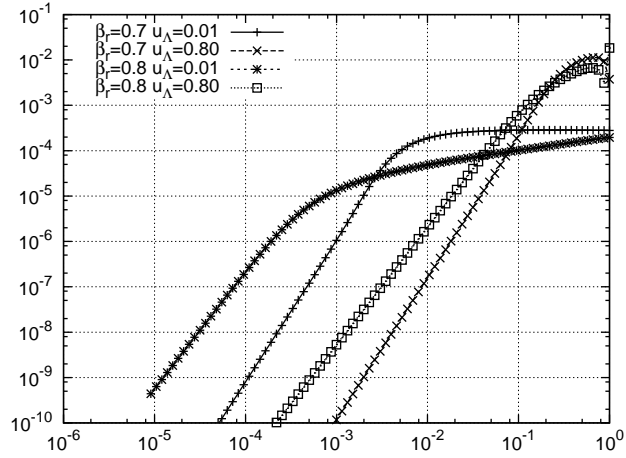}
\caption{The $\beta$-function of the curvature at the edge of the concave regions with respect to the scale $k$ near the fixed point for $\beta_r=0.7$ and $\beta_r=0.8$  for $\beta^2\tilde{u}_\Lambda=0.01$ and $\beta^2\tilde{u}_\Lambda=0.80$. This was computed with $b_p=2$.}\label{Power_strong_u1_impact_higher_beta}
\end{center}
\end{figure}

\begin{figure}[h!]
\begin{center}
\includegraphics[angle=270,scale=0.5]{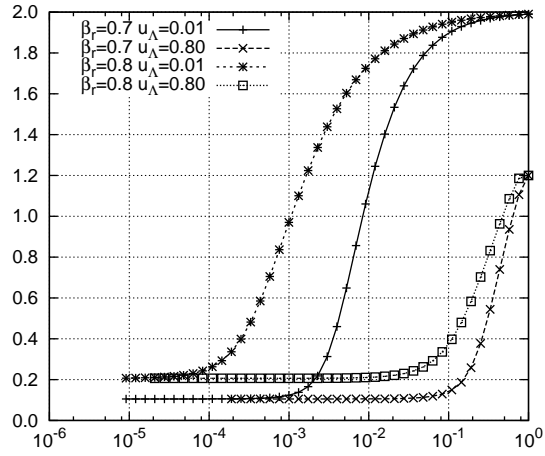}
\caption{The curvature $\ts''(\phi_n)$ with respect to the scale $k$ near the fixed point for $\beta_r=0.7$ and $\beta_r=0.8$ for $\beta^2\tilde{u}_\Lambda=0.01$ and $\beta^2\tilde{u}_\Lambda=0.80$. This was computed with $b_p=2$.}\label{Power_strong_u1_impact_higher}
\end{center}
\end{figure}

The universality found using Litim regulator is still present, namely the independence on the bare sine-Gordon amplitude of the IR $\beta$-functions, see fig. \ref{Power_strong_u1_impact_higher_beta}. 
The independence of the bare sine-Gordon amplitude is easily established as the curvature trajectory ends at the same value for the both weak and strong bare coupling, see figure \ref{Power_strong_u1_impact_higher}.

\begin{figure}[h!]
\begin{center}
\includegraphics[angle=270,scale=0.5]{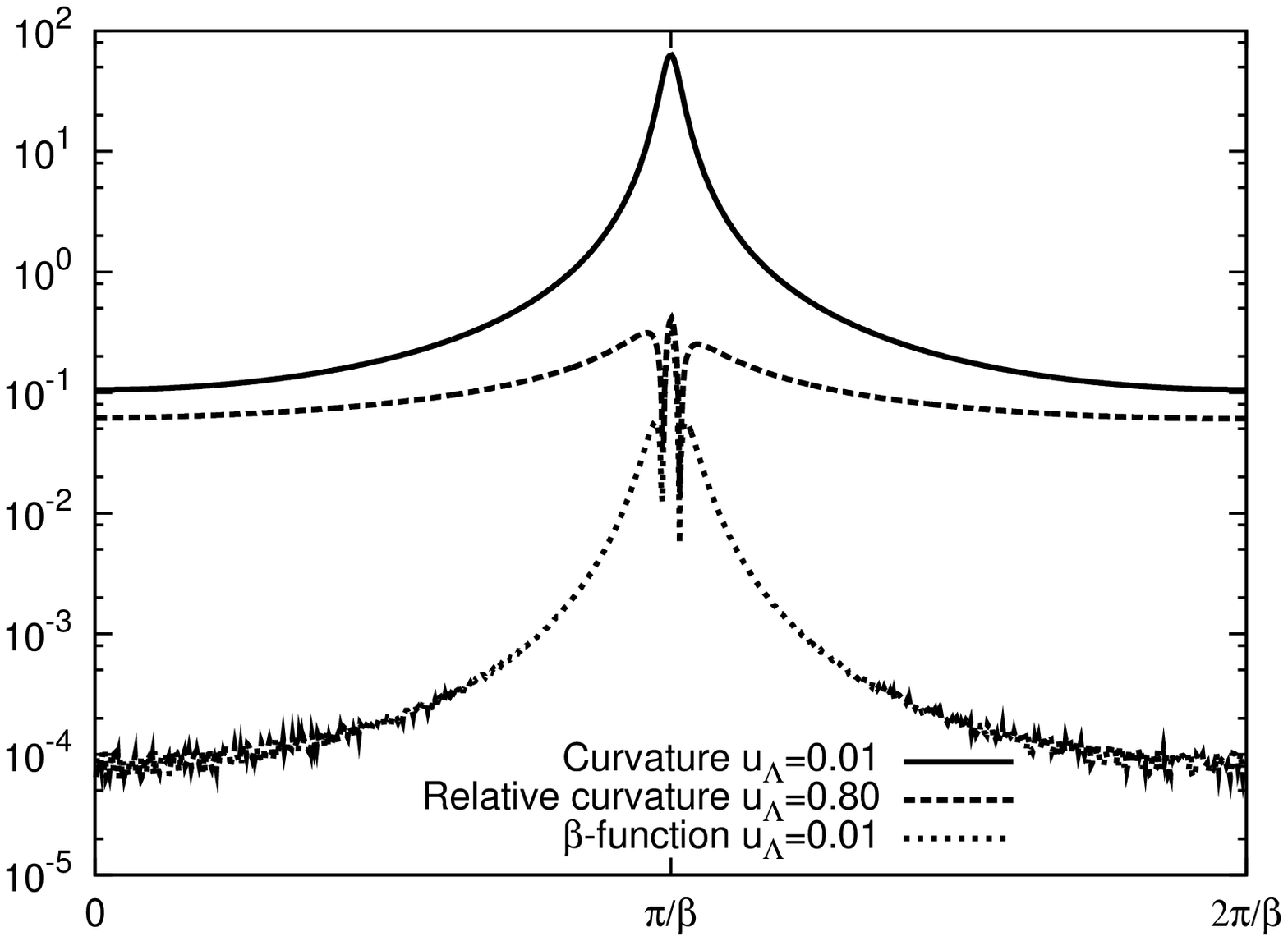}
\caption{From top to bottom  : the fixed point curvature $\ts''(\phi)$ for $\beta_r=0.7$ for $\beta^2\tilde{u}_\Lambda=0.01$ and the corresponding relative error times $10^3$ on the curvature computed for  $\beta^2\tilde{u}_\Lambda=0.80$. The last plot is the $\beta$-function for the whole curvature $|k\d_k\ts''(\phi)|$ times $10^5$ for $\beta^2u_{\Lambda}=0.01$. This was computed with $b_p=2$.}\label{Power_strong_u1_impact_relative}
\end{center}
\end{figure}

\begin{figure}[h!]
\begin{center}
\includegraphics[angle=270,scale=0.5]{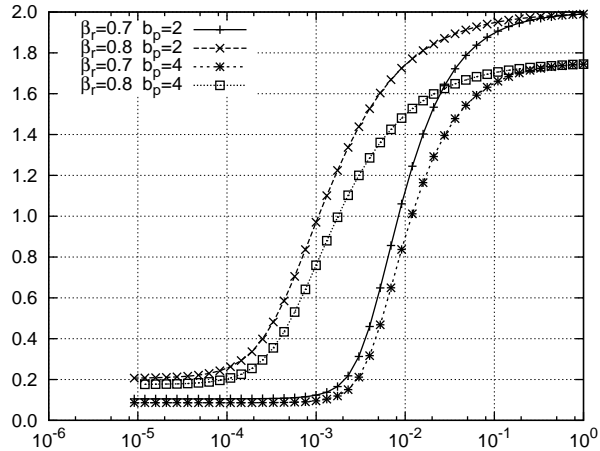}
\caption{The curvature $\ts''(\phi_n)$ with respect to the scale $k$ for $\beta_r=0.7$ and $\beta_r=0.8$ for $\beta^2\tilde{u}_\Lambda=0.01$ for different power-law regulators. }\label{Power_weak_u1_impact_higher}
\end{center}
\end{figure}

\begin{figure}[h!]
\begin{center}
\includegraphics[angle=270,scale=0.5]{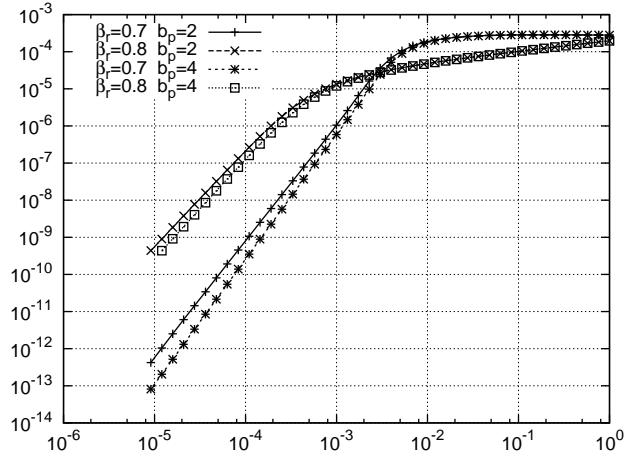}
\caption{The $\beta$-function of curvature $\ts''(\phi_n)$ with respect to the scale $k$ for $\beta_r=0.7$ and $\beta_r=0.8$ for $\beta^2\tilde{u}_\Lambda=0.01$ for different power-law regulators. }\label{Power_weak_u1_impact_higher_beta}
\end{center}
\end{figure}

As we saw previously, the whole curvature sees a set of fixed point that are the product of a finely tuned balance between the convex regions where the curvature is strongly enhanced and the concave ones where the curvature being attracted by the convexity limit tends to vanish, see figure \ref{Power_strong_u1_impact_relative}. 
At this stage, one can notice that the curvature is indeed the good order parameter as in this case $\tilde{k}^2=2$ but the curvature has still the same order of magnitude of values as in Litim's case. Obviously, the precise value depends on how the regulator suppresses the fluctuations. 
An obvious preoccupation is to study the dependence of the fixed-point solutions when changing the  regulator and in which cases we can get rid of this spurious effect.
Using parameterization (\ref{regulators}), one finds :
\be
\tilde{k}^2=\Big(a_p(b_p-1)\Big)^{\frac{1}{b_p}}\le[1+\frac{1}{b_p-1}\ri]
\ee This converges to 1 in the large $b_p$ limit slowly:
\be
\tilde{k}^2\simeq 1+\frac{log(b_p)}{b_p}
\ee and diverges in the large $a_p$ limit which amounts in rejecting the convexity singularity at $k=0$. We are obviously not interested in the latter limit as the IR fixed points are found at finite distance of this singularity. The figure \ref{Power_weak_u1_impact_higher} shows that the discrepancy between the fixed-point solutions for different power $b_p$ grows as we get closer to the Coleman fixed point. 
For the related $\beta$-functions, the behavior is the oppposite, as expected, see figure \ref{Power_weak_u1_impact_higher_beta}.

\begin{figure}[h!]
\begin{center}
\includegraphics[angle=270,scale=0.4]{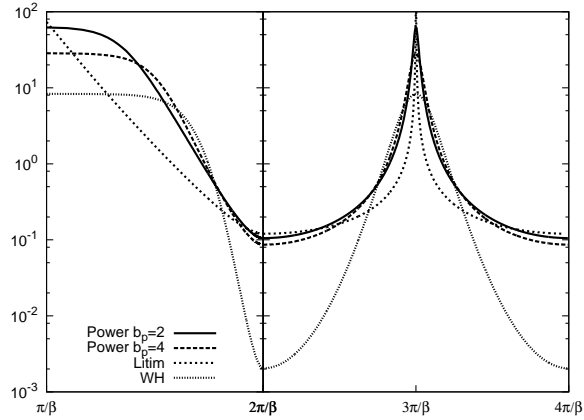}
\caption{The fixed-point curvature at the end of the flow with respect to $\phi$ for $\beta_r=0.7$ for $\beta^2\tilde{u}_\Lambda=0.01$ in $d=2$ for different regulators and Wegner-Houghton for comparison.}\label{Comp_shape}
\end{center}
\end{figure}

Finally, let us compare the shape of the fixed point curvature for the different RG schemes and regulators presented here, see figure \ref{Comp_shape}. In the convex regions, the curvature is strongly enhanced in all cases. While in Litim's case this amounts in an interaction that resembles a non-differentiable one, one can smoothly match the curvature given by average action with the one given by Wegner-Houghton when using the power-law regulator for increasing powers. This amounts in concluding that the would-be non-differentiability is not a genuine one, but rather an artifact of the non-analyticity of Litim regulator. In the concave regions, the difficulty of handling a system close to the limit of applicability of the loop-expansion for Wegner-Houghton is soften by the use of average action as its distance to the would-be singularity of the Legendre transform lies much further.

\section{Other dimensions}

Let us now focus on the case where the dimension is not 2. As we saw already, the dimensionfull flow equation preserves the periodicity in any dimension for all RG schemes.
 The dimensionfull effective potential can thus be Fourier-expanded at least in the UV perturbative regime :
\be
V_k(\phi)=\sum_{n=1}^{\infty}u_n(k)cos(n\beta \phi)
\ee where the dimensionfull frequency $\beta$ is a constant of the flow.
On the other hand, due to the finite scaling dimension of the field, the dimensionless flow equation breaks periodicity. This amounts in Wegner-Houghton RG to : 
\be
k\frac{\d \tv}{\d k}+d\tv-\frac{d-2}{2}\tp\tv'=-\frac{K_d}{2}Log(1+\tv'')
\ee
where the dimensionless effective potential can then be written in terms of dimensionless variables:
\be
\tv(\tp)=\sum_{n=1}^{\infty}\tu_n(k)cos(n\tb(k) \tp)
\ee
The dimensionless frequency is now scale dependent  \cite{Nandori:2003pk}  :
\be
\tb^2(k)=\beta^2k^{d-2}
\ee
The first obvious conclusion is that the Coleman frequency does not exist in $d\neq2$ as the dimensionless potential is not periodic : even if the amplitudes $\tu_n$ were seeing a fixed-point, the frequency would still run. As a result, any perturbative periodic bare theory is driven by the the gaussian fixed point.
Let us  linearize the flow equation in the UV weak coupling limit :
\be
k\frac{\d \tv}{\d k}+d\tv-\frac{d-2}{2}\tp\tv'+\frac{K_d}{2}\tv''=0
\ee 
Thus the UV perturbative flow is driven by :
\beq\label{higher_dim}
k\d_k \tu_n&=&\tu_nd\le(n^2\beta_r^2k^{d-2}-1\ri)\nn
\tu_n(k)&=&\tu_n(\Lambda)\le(\frac{\Lambda}{k}\ri)^de^{\frac{n^2\beta_r^2\Lambda^{d-2}}{d-2}\le(\le(\frac{k}{\Lambda}\ri)^{d-2}-1\ri)}
\eeq where we defined in a similar fashion as in $d=2$ the reduced frequency $\beta_r^2=\frac{K_d}{2d}\beta^2$. By abuse of notation we will note :
\be\label{Coleman_WH2}
\beta_c^{wh}=\sqrt{\frac{2d}{K_d}}
\ee as it reproduces the Coleman frequency in $d=2$ and separates two different UV behavior of the $\beta$-functions in $d\ne2$.
In the following, we use the UV cut-off $\Lambda=1$ for simplicity as it provides $\tb(\Lambda)=\beta$. 
\subsection{No symmetric phase in $d>2$}
Let us first study the situation for $d>2$. For $\beta_r<1/n$, the $\beta$-function of the mode $n$ is always negative and exhibits UV asymptotic freedom. This corresponds to the beginning of a broken phase scaling, where the coupling grows with the flow. At some point, the couplings are large enough to break the perturbative assumption needed to linearize the flow and the full flow equation has to be solved non-perturbatively to conclude about the continuation of the flow. 
 \begin{figure}[h!]
\begin{center}
\includegraphics[angle=270,scale=0.4]{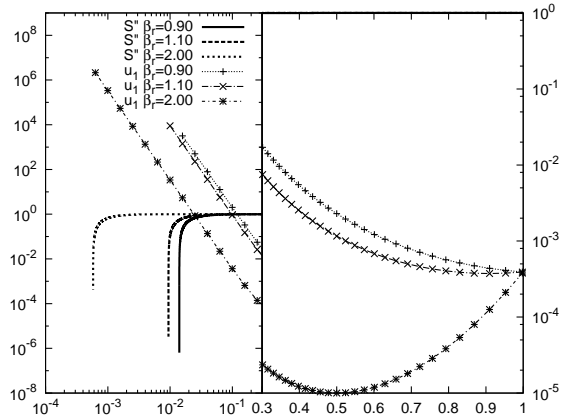}
\caption{The flow of the curvature $\ts''(\phi_n)$ and the first Fourier harmonic $\tu_1$ in units of $\beta^2$ with respect to the scale $k$ for different $\beta_r$ with $\beta^2\tu_1(\Lambda)=0.001$ in $d=4$. On the right hand side, the curvature value is given by $\approx 1$ due to the very weak bare coupling for the three different values of $\beta_r$.}\label{WH_flow_higher_amp_pot}
\end{center}
\end{figure}

For $\beta_r<1/n$, the $\beta$-function of the mode $n$ starts with a positive value and then changes sign at the scale $k_n=\le(\frac{1}{n^2\beta_r^2}\ri)^{\frac{1}{d-2}}$. In such a case, the trajectory drives the system closer and closer to the Gaussian fixed point as the frequency is increased to be finally reppeled to the broken phase in the IR. The UV flow of the curvature is then given only by the first harmonic as the higher ones, generated by the small non-linearites of the flow equation, are relatively strongly suppressed. These considerations are summarized in figure \ref{WH_flow_higher_amp_pot}. 
As a result, one can conclude that at LPA, the bare single harmonic sine-Gordon model is always driven by the perturbative flow to its broken phase. At the non-perturbative level, the IR behaviour of the figure \ref{WH_flow_higher_amp_pot} tells us that the broken phase behavior is not stabilized anymore by non-vanishing curvature fixed points as in $d=2$ and thus yields the appearance of the instability i. e. a degeneracy of the curvature signaling the breakdown of the loop-expansion. This instability, predicted to occur in \cite{Nandori:2003pk} using Fourier series, is demonstrated here to be a genuine one, being established with a global algorithm. 


\subsection{No broken phase in $d<2$}
\begin{figure}[h!]
\begin{center}
\includegraphics[angle=270,scale=0.4]{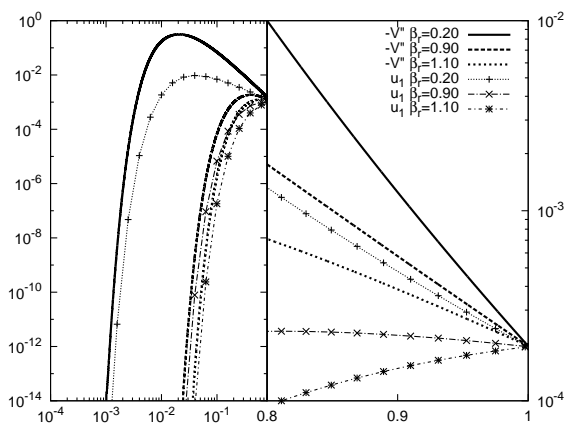}
\caption{The flow of the opposite of the potential curvature $-\tv''(\phi_n)=1-\ts''(\phi_n)$ and the first Fourier harmonic $\tu_1$ in units of $\beta^2$ with respect to the scale for different $\beta_r$ with $\beta^2\tu_1(\Lambda)=0.001$ in $d=1$. We see that starting with a small $\beta_r=0.2$, the trajectory goes in the non-perturbative sector $-\tv''(\phi_n)\simeq1$ but the IR scaling tends wash out the couplings.}\label{WH_flow_lower_amp_pot}
\end{center}
\end{figure}

The previous discussion still holds in the same fashion as before except that having $d<2$ reverses the behavior. More specifically, for $\beta_r<1/n$, the $n$-th sine-Gordon harmonic starts growing up to $k=k_n$ and then decreases. As a result, the perturbative scaling regime is shortened as the higher modes are enhanced and pile up during the evolution. Provided that $\beta_r$ is not too small, one can nevertheless preserve a perturbative trajectory that will at the end of the flow suppress all the couplings. For $\beta_r<1/n$, the flow monotonously suppresses the couplings. These considerations are illustrated on figure \ref{WH_flow_lower_amp_pot} where we show that even when the flow reaches the non-perturbative neighborhood of the instability, it finally goes toward the gaussian fixed point.

\subsection{Adding a regulator}
\begin{figure}[h!]
\begin{center}
\includegraphics[scale=1.2]{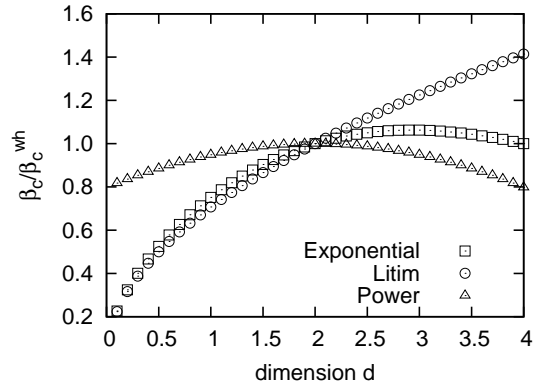}
\caption{The $\beta_c$ frequency counted in units of the Wegner-Houghton $\beta_c^{wh}$ frequency (\ref{Coleman_WH2}) against the dimension of the spacetime considered for various regulators for the parameters $a_e=b_e=c_e=1$ and $a_p=1, b_p=2$, cf .eq (\ref{regulators}), (\ref{AA_beta}) and (\ref{AA_exp}). Note the common crossing at $d=2$.}\label{beta_comp}
\end{center}
\end{figure}

Let us finally discuss how to adapt the results in dimension $d\ne2$ when using average action. Obviously, one will linearize the flow equation the same way in the UV but the loop-integral makes the evaluation of the prefactor of $\tv''$ more involved to establish. The only change is then the definition of $\beta_r$ and $\beta_c$ which are now given respectively for Litim, power-law ($a_p=1$) and exponential ($a_e=1$) regulators by :
\beq\label{AA_beta}
\beta_c^l&=&\sqrt{d/2}\beta_c^{wh}\nn
\beta_c^p&=&sinc\le(\pi\frac{d-2}{2b_p}\ri)^{1/2}\beta_c^{wh}\nn
\beta_c^e&=&\le(b_e^{-\frac{d}{2}+1}\Gamma(\frac{d}{2c_e}+\frac{c_e-1}{c_e})\ri)^{\frac{-1}{2}}\beta_c^{wh}
\eeq where the exponential regulator is parametrized by:
\beq\label{AA_exp}
r_e(y)&=&\frac{a_e}{e^{b_ey^{c_e}}},\quad (a_e,b_e,c_e)\in (\R^{+*})^2\times[1:\infty]
\eeq
The behaviour of the $\beta_c$ frequency of different regulators with respect to a change of dimension  is summarized in figure \ref{beta_comp}. 
It has been numerically tested that the features obtained by Wegner-Houghton RG can be qualitatively  reproduced, namely the presence of the instability in the IR for $d>2$ and the rebounce driving to the gaussian FP in $d<2$.
\section{Discussion}\label{sect5}
\subsection{Universality in the broken phase for $d=2$}
In this paper, we established that at LPA using Wegner-Houghton RG, the sine-Gordon model exhibits a continuous line of non-perturbative IR fixed points in $d=2$. It was shown that these fixed-points could not be found using Fourier expansion, suggesting that the relevance classification around these new fixed-points may not be given by the usual simple UV relevant trigonometric functions. It was also demonstrated that the fixed points are independent of the bare sine-Gordon amplitude and the universality of the $\beta$-functions in the IR has been demonstrated. 

The set of fixed points found in the broken phase allowed us to define the critical exponent for the susceptibility in this model $\gamma\simeq1$ for Wegner-Houghton when fitted for $\beta_r>0.95$ from figure \ref{Litim_phase_summary}. Obviously, this value of critical exponent is only of little help as it was computed near $\beta_r=1$ where the wave-function renormalization effect is expected to be large. The next order of the gradient expansion will be addressed in another paper \cite{Pangon:2010sg}.
\subsection{The convexity issue and the Fourier convergence}
As it is well known, the IR behavior of a scalar theory in its broken phase is usually driven by a dynamical Maxwell-cut to recover the proper convexity property. 
Its simplest realization is given by the effective potential $\tv_k=-\frac{1}{2}\tp^2$ in the concave regions. As it has been detailed for both Wegner-Houghton and Average action RG, this approach to convexity is usually identified as a tree-level singularity for the flow equation. We showed here that in $d=2$  the loop-evolution can mimic this convexity approach consistently, without sending the flow into a singular region where the flow equation does not hold anymore at least in a part of the broken phase $\beta_r>0.55$, the action being stabilized by IR fixed points. Due to the large degenerated sectors of the effective action, the Fourier serie was shown to exhibit a very shallow convergence. In $d>2$, the Fourier convergence is obviously still a problem close to the convexity limit. Using a global resolution not relying on Fourier analysis, we demonstrated that the IR flow may run into a genuine singularity. The absence of stabilization mechanism can be explained in two ways.

First, the dimension 2 is a very particular case as it allows to have stable kinks in the broken phase. These inhomogeneous configurations contributions to the flow may lend some stability to the IR behavior. The lack of proof of stabilization of the flow for too small $\beta_r$ can be seen whether as a too weak numerical accuracy whether as the presence of new zero modes : it is known that for low enough $\beta_r$, kink bound-states appear in the spectrum.

Another possible explanation is the non-vanishing scaling dimension in $d\ne2$. In this case, the dimensionless $\tb$ frequency decreasing with the flow for $d>2$, the size of the concave regions in the space of the dimensionless field configurations is growing as $\frac{2\pi}{\tb}$. As a result, this enforces the degeneracy of the action in the corresponding regions when the scale is decreased. On the opposite, in $d<2$, the size of the field configuration space where the action is not convex decreases with the flow, and the convexity limit that we could find starting with $\beta_r<<1$ is expected  to be reppelling.

\subsection{The mass gap}
As widely known from semi-classical considerations, the broken phase of the sine-Gordon model exhibits a mass gap due to kinks and kink bound states. The set of fixed points found here are clearly incompatible with it as this would mean that the dimensionful curvature around the minima is going to zero when integrating all the fluctuations. 
This is the price to pay for the use of LPA : the kink mass which is usually recovered by RG to be an invariant of the perturbative flow with wave function renormalization crucially depends on the Kosterlitz-Thouless separatrices presence. Nevertheless, we saw that even at LPA, one has some traces of the mass of the excitations : the convex regions were strongly enhanced in the IR before saturating. 

As a result, we can not hope for a global fixed point for the action when including  wave-function renormalization non-perturbatively, as the convex regions are expected to exhibit a mass gap.

\subsection{The use of average action}
Another aspect of the present study was to clarify the impact of the regulator when using average action as its introduction breaks periodicity. It was shown that for any regulator, the phase boundary does not change in $d=2$. Focusing on the broken phase, we exhibited a similar set of fixed points in the IR that have been shown to be bare-amplitude independent for various regulators. This feature showed that even if the regulator changes quantitatively the broken phase -the critical exponent is $\gamma\simeq0.4$ from the fit for $\beta_r>0.9$ using Litim regulator- it allows to reproduce the main features qualitatively. In other dimensions, even if the frequency $\beta_c$ is different as the one in Wegner-Houghton, one can demonstrate that the results are the same provided the use of the relative frequency $\beta_r$.

In $d=2$, as the broken phase appeared to be involved, being the product of a fine balance between convex and concave regions of the potential to stabilize the potential above the flow singularity, it can also be seen as an ideal testing ground for approximate RG schemes, namely proper-time. In \cite{Bonanno:2004pq,Consoli:2006ji,Litim:2006nn}, proper-time flows where used to study the broken phase of $\phi^4$ model where it was argued that there was no conceptual problem in reaching the limit $k=0$. It would be interesting to check if the fixed-points are still present, preventing the appearance of a singularity in the flow and if the convex region is always well-behaved. Clearly, for the "exponential case" of proper-time, that was found to optimize the computation of critical exponents in \cite{Mazza:2001bp},  there is little hope to find the fixed points being found here at a finite distance of the convexity limit that is in this case rejected at $k=0$.

As mentionned in the introduction, the average action has been extensively used to study Yang-Mills theories deconfinement transition. Due to the presence of the regulator breaking gauge invariance, one can use e. g. modified Ward identities to wash out its effects \cite{Gies:2006wv}. The problem of such process is that the center symmetry breakdown that is monitored to study deconfinement is clearly unconstrained by such procedure, being a discrete symmetry. Our present finding that the addition of a regulator is qualitatively priceless is the case of the sine-Gordon discrete periodicity symmetry breaking  conforts the use of this approach.

\subsection{The phase structure for $d\neq 2$}
Finally, even simple perturbative considerations allowed us to demonstrate that starting with a perturbative sine-Gordon model always drives us in the broken phase for $d>2$. The absence of a symmetric phase at LPA order is not really good news if we are to mimic $SU(2)$ with the sine-Gordon model, as it would correspond to the absence of a confining phase. 

This pushes us to conclude that the Haaron gas is more involved than a simple perturbative sine-Gordon and that the higher order harmonics dropped for simplicity in \cite{Johnson:1991qc,Polonyi:1995jz} have to be taken into account in $d>2$ due to the absence of Coleman fixed point needed to use universality argument on the first harmonics. 
In $d<2$, the symmetry seems to be always preserved, reflecting the fact that $d=2$ is the lower critical dimension.

In any case, we saw that the main features in $d\ne2$ are driven by the scaling dimension of the field so that these conclusions can perfectly be not representative of the genuine behavior when wave-function renormalization is taken into account, especially in the non-perturbative sectors.

\section*{acknowledgments}
The author wishes to thank J. Polonyi for many fruitful discussions and J. P. Blaizot, B. Delamotte and D. Mouhanna for their constructive suggestions in the improvement of the manuscript.
The author is also grateful to B. Friman, V. Skokov and S. Nagy for many useful advices. 
\appendix
\section{Numerical details}\label{app_num}
The numerical study of flow equations is in general an involved problem as we study a strongly non-linear partial differential equation (PDE) or an integro-partial differential equation (IPDE) when the loop-integral is analytically intractable. Many studies made the use of the algorithm of \cite{Adams:1995cv} using finite differences in a grid of $\phi$. This method has been shown to be very powerful, especially in computing critical exponents \cite{Berges2002}.

We made use in the present study of the algorithm developed by Berzins and Dew \cite{BerzinsDew0}, \cite{BerzinsDew1} and \cite{BerzinsDew2} known as $C^0$ collocation  method. The main reason is because it is much faster when dealing with a very fine dependence on $\phi$ as meshes of $10^4$ points can be computed on a laptop within a reasonable amount of time. It has been shown to reproduce accurately the analytical results \cite{Litim:2002cf} for the Wilson-fixed point potential in $d=3$ in \cite{Pangon:2009pj}.
The algorithm relies on two different parts. The first one, the most important in our case, is the discretization of the field dependence using spline representation i.e. Chebyshev piecewise polynomials e. g. \cite{Diaz1977} and \cite{10705}. This parameterization exhibits a very fast convergence with respect to two parameters : the degree of the polynomials chosen on each sub-interval and the number of sub-intervals into which the field interval is spitted into. \\
Due to the finiteness of the range of $\phi$ for which the flow equation is solved, two boundary conditions are needed. In the special case of sine-Gordon model, we constrained the first derivative of the curvature to vanish in $\phi=0$ due to the manifest $Z_2$-invariance of the problem. As we solved the flow on an integer number of periods $\frac{2\pi}{\beta}$, the other boundary has been chosen to follow the same condition due to periodicity. Even if periodicity was not explicitly encoded  in our numerical solution, we found that solving the flow equations over $10$ periods were enough to get rid of any finite boundary effect.\\
The field discretization was dealt on $2000$ sub-intervals, each one containing a Chebyshev polynomial of degree $3$. For low $\beta$ -when the points density is smaller- increasing ten times the number of points as been tested to exhibit the same results in Wegner-Houghton case and with the power-law regulators. In the case of Litim regulator, we tested systematically the numerical stability against an increase of number of points and even when the results slightly changed, it never excessed $10^{-3}$ in relative error.\\
The scale integration is handled by the DASSL routine \cite{Brenan1989} using backward differentiation formula that is known to be efficient in estimating the scale derivative in stiff cases.

Finally, the loop-integrals for the power-law regulator are performed using adaptive Gauss-Kronrod rules. Even if the explicit formulation of the flow equation can be worked out, cf appendix \ref{app_power}, we chose to perform numerically the integration. First because the use of the analytical formulation does not save much computation time due to the appearance of non-trivial functions (mostly the $arctan$) but also because we were not able to compute the explicit shape of the flow equation for other power-law regulators. As we compared the different flows for $b_p=2$ and $b_p=4$, we wanted to have the same degree of accuracy in both cases.  Let us also mention that in the case of the sine-Gordon model, a fit of the loop integral as a function of the curvature is not a well suited strategy as this function would have to be fitted for the whole range of the possible curvatures. In the concave regions, we saw that it gets close to an instability while in the convex regions, a very accurate estimate is needed for large curvature values.

\section{Callan-Symanzik RG}\label{app_internal}
Let us briefly present the so-called Callan-Symanzik RG  \cite{Litim:1998nf,Alexandre:2000eg,Alexandre:2001wj}. 
In this scheme, the running quantity is not a coarse graining scale but a bare mass and thus the flow equation describes how different bare theories are related through a change of the mass and reads :
\be
k\d_k\Gamma[\phi]=Tr\le(\frac{k^2}{k^2+\Gamma^{(2)}}\ri)
\ee where the functional kernel of the right-hand side have been demonstrated to be related to the proper-time flows generators, \cite{Litim:2002xm}.
In general, this flow equation still contains UV (and IR) divergences as no explicit regularization has been made. Let us now imagine that we are able to compute the renormalized action for some well-chosen value of $k$. This flow equation is now nothing than the PDE one will have to solve to transport the known renormalized result (usually perturbative for $k$ large enough) to a region of physical interest (typically $k=0$). Obviously, this procedure will work provided that a continuous mapping can be made between the bare and the renormalized parameters i.e. provided that no phase boundary were encountered.

Let us now apply this equation on the simple ansatz :
\be
\Gamma_k[\phi]=\int d^dx \le[\frac{z_k}{2}\d_\mu\phi\d_\mu\phi+V_k[\phi]\ri]
\ee where $z_k$ is field independent. The flow equation for the potential reads :
\be
k\d_kV_k=K_d\int p^{d-1}dp \frac{k^2}{k^2+z_kp^2+V_k''(\phi)}
\ee
that can be derived from the power-law regulator for $a_p=1, b_p=1$ using average action. This integral is straightforward computed in $d=1$ as it is well known that no divergences occur in Quantum Mechanics. In $d=2$, the integral exhibits a logarithmic divergence that should be subtracted :
\beq
k\d_kV_k
&=&K_2\int pdp \frac{k^2}{z_kp^2+k^2}\sum_{n=0}^{\infty}(-1)^n\le(\frac{V_k''(\phi)}{z_kp^2+k^2}\ri)^n\nn
&=&-\frac{K_2}{2}\frac{k^2}{z_k}Log\le(1+\frac{V''}{k^2}\ri)
\eeq where the term $n=0$ was dropped being an infinite field -independent constant. This equation is nothing but the Wegner-Houghton equation with a field-independent wave function renormalization. At this stage, one can remember the perturbative proof of \cite{Amit:1979ab} to establish the perturbative renormalization group equations : starting with a massive sine-Gordon model, all the computations are well defined and the vanishing mass limit is taken only at the end. So that this procedure can be seen as a perturbative version of Callan-Symanzyk RG.
Non-perturbatively, we know this procedure does not meet a singularity on the trajectory as the line of fixed points we established for Wegner-Houghton holds and protects the evolution. 
As a result, we believe this provides a non-perturbative proof of the consistency of the approach of \cite{Amit:1979ab} that has to be checked otherwise only order by order in the perturbative expansion. It also shows that no-phase boundary is encountered when smoothly switching off the mass term in the massive sine-Gordon model.

Finally, let us emphasize that the flow equation have been safely regularized by the possibility of dropping an infinite contribution to the free-energy that was in our ansatz field-independent. If we were to take into account a field-dependent wave-function renormalization, it is straightforward to notice that this simple recipe does not apply, not even considering the flow equation for $z_k$. As a result, it may be the sign that perturbative renormalization procedure \cite{Amit:1979ab} is lost if we were to study more involved ansatz.

\section{Power-law regulator}\label{app_power}
While in general, the loop-integral can not be computed in the flow equation of average action, it is the case for the power-law regulator for $a_p=1$ and $b_p=2$. The first point of interest is that it emphasizes the special role of the quantity $\tilde{k}$ as the flow equation has two different analytical expressions that are continuously matching at $\tv''=\tilde{k}^2$ in every dimension. 
The equation (\ref{AA_flow_dimless}) reads in dimension 2 \cite{Nandori:2009ad}:
\beq
&&k\d_k \tv +2\tv\nn 
&\stackrel{\tv''>2}{=}&-\frac{K_2}{\sqrt{\tv''^2-4}}Log\le(\frac{\tv''-\sqrt{\tv''^2-4}}{\tv''+\sqrt{\tv''^2-4}}\ri)\nn
&\stackrel{|\tv''|<2}{=}&-\frac{K_2}{\sqrt{4-\tv''^2}}\le[2arctan\le(\frac{\tv''}{4-\tv''^2}\ri)-\pi\ri]\nn
\eeq Let us emphasize that at $\tv''=2$, at the junction of the two analytical expressions, this expression is not only continuous in $\tv''$ with a common value $\frac{8K_2}{3}$ but also infinitely many differentiable. This property of smooth transition between the two regions is preserved in the other dimensions as well.
In dimension 3, the flow equation reads :
\beq
&&k\d_k \tv +3\tv-\frac{\tp\tv'}{2} \nn
&\stackrel{\tv''>2}{=}&-K_3\frac{\pi}{\sqrt{2}}
\Bigg[
\frac{\tv''}{\sqrt{\tv''^2-4}}\Big(
  \frac{1}{\sqrt{\tv''-\sqrt{\tv''^2-4}}}\nn
&-&\frac{1}{\sqrt{\tv''+\sqrt{\tv''^2-4}}}  \Big)\nn
&-&\frac{1}{\sqrt{\tv''-\sqrt{\tv''^2-4}}}-\frac{1}{\sqrt{\tv''+\sqrt{\tv''^2-4}}}   
\Bigg]\nn
&\stackrel{|\tv''|<2}{=}&K_3\frac{\pi}{\sqrt{\tv''+2}}
\eeq
Finally, the situation in dimension 4 is a little more complicated as the flow equation as a superficial divergence in the large momentum limit. However, this divergent term is field-independent and can be safely subtracted as performed in appendix \ref{app_internal}. Unfortunately, after this subtraction, the remaining can not be recasted easily in terms of elementary functions so that the analytical form of the flow equation is rather given for $\tv'$:
\beq
&&k\d_k \tv' +4\tv' -\tp\tv'' \nn
&\stackrel{\tv''>2}{=}&-\frac{K_4\tv'''}{(\tv''^2-4)^{3/2}}\Bigg[\tv''\sqrt{\tv''^2-4}\nn
&+&2Log\le(\frac{\tv''-\sqrt{\tv''^2-4}}{\tv''+\sqrt{\tv''^2-4}}\ri)\Bigg]\nn
&\stackrel{|\tv''|<2}{=}&-\frac{2K_4\tv'''}{(4-\tv''^2)^{3/2}} \Bigg[2arctan\le(\frac{\tv''}{\sqrt{4-\tv''^2}}\ri)\nn
&-&\pi-\sqrt{4-\tv''^2}\Bigg]
\eeq

 \bibliographystyle{plain}	


\end{document}